\documentclass[%
 reprint,
relation of the surface states 
 amsmath,amssymb,
 aps,
 prb,
dvipdfmx
]{revtex4-1}
\usepackage[dvipdfmx]{graphicx}
\usepackage[dvipdfmx]{color}
\usepackage{dcolumn}

\usepackage{amsmath} \usepackage{amssymb}
\usepackage{accents} \usepackage{bbm}
\usepackage{mathrsfs}
\usepackage{accents}
\usepackage{multirow}

\newcommand{\bs}[1]{\boldsymbol{#1}}

\newcommand{\gns}[0]{G_{\mathrm{NS}}}
\newcommand{\gnn}[0]{G_{\mathrm{NN}}}

\newcommand{\ua}[0]{\uparrow} \newcommand{\da}[0]{\downarrow} 

\definecolor{dgreen}{RGB}{00, 120, 00}
\definecolor{dgreen}{RGB}{00, 00, 120 }
\usepackage{hyperref}
\hypersetup{colorlinks=true, linkcolor=blue, urlcolor=blue, citecolor=blue, }
\usepackage{booktabs}

\usepackage{ulem}

\begin{document}
\preprint{APS/123-QED}

\title{Identifying possible pairing states in Sr$_2$RuO$_4$ by tunneling spectroscopy}

\author{Shu-Ichiro Suzuki$^{1}$}
\author{Masatoshi Sato$^{2}$}
\author{Yukio Tanaka$^{1}$}

\affiliation{$^{1}$Department of Applied Physics, Nagoya University, Nagoya 464-8603, Japan}
\affiliation{$^{2}$Yukawa Institute for Theoretical Physics, Kyoto University, Kyoto 606-8502, Japan}

\date{\today}

\begin{abstract}
We examine the tunneling spectroscopy of three-dimensional
normal-metal/Sr$_2$RuO$_4$ junctions as an experimental means to
identify pairing symmetry in Sr$_2$RuO$_4$. In particular, we consider
three different possible pairing states in Sr$_2$RuO$_4$: spin-singlet
chiral $d$-wave, spin-triplet helical $p$-wave, and spin-nematic
$f$-wave ones, all of which are consistent with recent
nuclear-magnetic-resonance experiments [A. Pustogow \textit{et al.},
\href{https://doi.org/10.1038/s41586-019-1596-2}{Nature \textbf{574},
72 (2019)}].  The Blonder-Tinkham-Klapwijk theory is employed to
calculate the tunneling conductance, and the cylindrical
two-dimensional Fermi surface of Sr$_2$RuO$_4$ is properly taken into
account as an anisotropic effective mass and a cutoff in the momentum
integration.  It is pointed out that the chiral $d$-wave pairing state
is inconsistent with previous tunneling conductance experiments along
the $c$-axis.  We also find that the remaining candidates, the
spin-triplet helical $p$-wave pairing state and the spin-nematic
$f$-wave ones, can be distinguished from each other by the in-plane
tunneling spectroscopy along the $a$- and $b$-axes.
\end{abstract}

\pacs{???}
\maketitle



\section{Introduction}
The pairing symmetry of Sr$_2$RuO$_4$ has been a mystery since its
discovery\cite{Maeno_Nature_1994, Mackenzie_RMP_2003,
Maeno_JPSJ_2012}.  Until recently, the most promising candidate had
been the chiral $p$-wave (i.e., $p_x + i p_y$-wave) pairing
\cite{Rice_1995}, which is spin-triplet with broken time-reversal
symmetry (TRS). 
The spin-triplet pairing was widely accepted since it is consistent
with a variety of experiments such as
polarized-neutron-scatterings\cite{Duffy_PRL_2000}, half-quantum
vortices\cite{Jang_Science_2011,Yasui_PRB_2017}, the transport
measurements \cite{Laube_PRL_2000, Kashiwaya_PRL_2011, Yamashiro_JPSJ_1998, 
Jin_PRB_1999, TYBN09, Samokhin, Anwar_NatComn_2016,
OldeOlthof_PRB_2018}, and in particular the nuclear-magnetic-resonance
(NMR) measurement\cite{Ishida_Nature_1998}.  There were also a number
of theoretical studies supporting the spin-triplet
scenario\cite{Nomura, Nomura08, Yanase, Raghu, Kohmoto, Kuroki,
Takimoto,Tsuchiizu}.  However, the situation has been changed after a
recent report \cite{Pustogow_arXiv_2019} pointing out an over-heating
problem in the previous NMR measurements. The new NMR data without the
heating problem show the reduction of the in-plane spin susceptibility
\cite{Pustogow_arXiv_2019, Ishida_arXiv_2019} below $T_{\rm c}$, which
conflicts with the in-plane equal spin structure of the chiral
$p$-wave pairing where the $d$-vector is pinned along the
$c$-axis.  Moreover, a first-order phase transition of the
superconducting state triggered by an in-plane magnetic field
\cite{Yonezawa_PRL_2013, yonezawa_JPSJ_2014, Kittaka_PRB_2014} also
suggests a different spin structure.

In addition to the spin structure, TRS in the superconducting state
has been controversial.  The spontaneous TRS breaking has been
reported by the muon-spin-relaxation ($\mu$SR) and the Kerr-effect
measurements\cite{Luke_Nature_1998, Xia_PRL_2006}.  The ultrasound
measurements \cite{Lupien_PRL_2001, Okuda_JPSJ_2002} also suggest a
two-dimensional (2D) gap function, which is also consistent with 
broken TRS.  In contrast, the spontaneous edge current associated with
a chiral state has never been observed so far
\cite{Matsumoto_JPSJ_1999, Furusaki_PRB_2001, Bjornsson_PRB_2005,
Kirtley_PRB_2007, Hicks_PRB_2010, Kallin_2012, Curran_PRB_2014, SIS3,
Bakurskiy_2017}.  A recent report on the Josephson effects also
supports the presence of TRS \cite{Kawai_PRB_2017, Kashiwaya2019}.
Moreover, two different nodal structures of the superconducting gap
have been reported.  The thermal-conductivity
\cite{Hassinger_PRX_2017} and specific-heat measurements
\cite{Kittaka_JPSJ_2018} suggest a vertical and horizontal line node
(or gap minimum), respectively. 

Two alternative pairing states have been proposed to explain the
Knight-shift measurements\cite{Pustogow_arXiv_2019,
Ishida_arXiv_2019}, together with a part of the other experiments.
One is a chiral $d$-wave (i.e., $d_{zx} + i d_{yz}$-wave) pairing
\cite{Zutic_PRL_2005}, which is spin-singlet with broken TRS and a
horizontal line node.  The other is a helical $p$-wave pairing state
that preserves TRS and may reproduce the in-plane transport
measurements \cite{Kashiwaya_PRL_2011}.  The latter pairing is
an one-dimensional irreducible representation and has no node on the
Fermi surface (FS). 

We also would like to point out here that spin-nematic pairings could
be consistent with several experiments including the recent NMR data:
They are spin-triplet and compatible with the NMR experiments if their
$d$-vector points to the direction of the applied magnetic fields
(i.e., the $a$-axis).  They can also be consistent with the
ultrasound measurements and the Josephson effects since the
spin-nematic pairings are multi-dimensional with TRS.  Moreover, a
spin-nematic $f_{xyz}$-wave pairing [see Eq.~\eqref{eq:f-wave}] can
reproduce the four-fold symmetric superconducting gap with horizontal
and line nodes.   

Although these pairing states are not fully consistent with all of the
existing experiments, all of them may reproduce the NMR Knight-shift
measurements. Obviously, a solid experimental means that can
distinguish the above pairings is highly desired. 

\begin{figure}[b]
	\centering
  \includegraphics[width=0.46\textwidth]{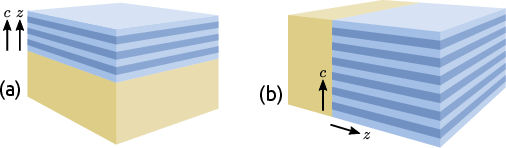}
	\caption{Schematics of the three-dimensional junctions. 
	The stripes represent the layers of RuO$_2$ planes. 
	The $c$-axis of Sr$_2$RuO$_4$  is (a) parallel and (b) perpendicular to the
	interface normal. }
	\label{fig:Sche}
\end{figure}
A remarkable property of unconventional superconductors (SCs)
including Sr$_2$RuO$_4$ is the
presence of surface Andreev bound states\cite{Hara_PTP_1986} (ABSs).
The bound states are formed at a boundary of an SC when the phases of
the pair potential for incoming and outgoing quasiparticles are
different.  \cite{Bruder,TK95,ABS,Yamashiro97, Honerkamp98, Hu, ABSR1,
ABSR2, Asano_PRB_2004}  Their energy dispersion reflects the internal
phase of the anisotropic pairing, and manifests as  a zero-energy peak
(ZEP) in the conductance spectra.  A sharp ZEP appears when the ABSs
form a flat band, while a dome-shaped broad ZEP arises when the ABSs
are dispersive \cite{TK95, ABS, Yamashiro97,TYBN09}.  Comparing the
conductance spectra in details, one can obtain the information of the
pair potential with which the above three pairing states can be
distinguished. 

In this paper, we propose the tunneling spectroscopy of
three-dimensional normal-metal/Sr$_2$RuO$_4$ junctions as an
experimental means to determine the pairing symmetry of Sr$_2$RuO$_4$.  
We examine three-dimensional junctions as shown in Fig.~\ref{fig:Sche}
where the cylindrical FS of Sr$_2$RuO$_4$ (see Fig.~\ref{fig:Sche_FS})
is taken into account. We consider the spin-singlet $d_{zx} +
id_{yz}$-wave, spin-nematic $f_{xyz}$-wave, and spin-triplet helical
$p$-wave pairing states.  It is shown that the $d_{zx} +id_{yz}$-wave
pairing state hosts a sharp robust ZEP at the (001) surface.
Comparing it with the spectroscopy data by the scanning tunneling
microscope (STM) \cite{Suderow_2009, Firmo_PRB_2013, Madhavan}, we exclude the
$d_{zx}+id_{yz}$ pairing from possible pairings of Sr$_2$RuO$_4$.
Even though a simple spin nematic $f_{xyz}$-wave pairing has a similar
ZEP at the (001) surface, this peak is fragile and easily suppressed
by, for example, the Rashba spin-orbit interaction (RSOI) at the
interface.  Thus, the spin-nematic state could be consistent with the
STM data.  Then the spin-triplet helical $p$-wave pairing naturally
reproduces the STM data.  The latter two pairings, spin-triplet
$f_{xyz}$-wave and helical $p$-wave pairings, can be distinguished by
the conductance spectra of the (100)- and (110)-interface junctions. For the
$f_{xyz}$-wave junctions,  the conductance spectra are different
between these junctions: the ZEP appears in the (100) case but a
V-shaped spectrum dose in the (110) case. In contrast, those for the
helical $p$-wave SC are qualitatively identical. 

\section{Blonder-Tinkham-Klapwijk theory}
In this paper, we consider three-dimensional junctions as shown in Fig.~\ref{fig:Sche}. A 
normal metal (N) and an SC occupy $z<0$ and $z \geq 0$, respectively. The
junction is assumed infinitely large in the $x$ and $y$ directions. 
The interface normal vector $\bs{e}_z$ is perpendicular or parallel to
the $c$ axis of Sr$_2$RuO$_4$ as shown in Figs.~\ref{fig:Sche}(a) and
\ref{fig:Sche}(b), respectively.

The Hamiltonian for superconducting systems is given by 
\begin{align}
  &\mathcal{H} 
	= \frac{1}{2} \, \int 
	\Psi^\dagger (\bs{r})
	\check{H}_{B}(\bs{r}) 
	\Psi         (\bs{r})
	\, d\bs{r}, 
	\\[0mm]
	&\Psi           (\bs{r}) =
	[~
   \psi_{\uparrow}           (\bs{r}) \hspace{2mm} 
	 \psi_{\downarrow}         (\bs{r}) \hspace{2mm} 
	 \psi_{\uparrow}  ^\dagger (\bs{r}) \hspace{2mm} 
   \psi_{\downarrow}^\dagger (\bs{r}) ~]^T, 
\end{align}
with the Bogoliubov-de Gennes (BdG) Hamiltonian 
\begin{align}
	& \check{H}_{B}(\bs{r}) =
	\left[
		\begin{array}{cc}
			 \hat{h     }  ( \bs{r}) &
			 \hat{\Delta}  ( \bs{r}) \\[2mm]
			-\hat{\Delta}^*( \bs{r}) &
			-\hat{h     }^*( \bs{r}) \\
	\end{array}\right], 
	\label{eq:BdG}
	\\[0mm]
	& \hat{\Delta}  ( \bs{r}) =
		i \left[ d_0(\bs{r}) + \bs{d}(\bs{r}) 
    \cdot \hat{\bs{\sigma}} \right] \hat{\sigma}_y
    \Theta(z), 
\end{align}
where $d_0$ and $\bs{d}$ are the spin-singlet and spin-triplet
components of the pair potential, $\hat{\sigma}_0$ and
$\hat{\sigma}_\nu$ ($\nu={x,y,z}$) are
the identity and the Pauli matrices in the spin space, and $T$ being the transpose of a
matrix.
Throughout this paper, the symbol
$\hat{\cdot}$ ($\check{\cdot}$) represents a $2 \times 2$ ($4 \times 4$)
matrix in the spin (spin-Nambu) space. 

The single-particle Hamiltonian $\hat{h} (\bs{r})  
= 
	 \hat{h}_{\mathrm{SP}} 
	+\hat{h}_{\mathrm{SO}} 
	+\hat{h}_{\mathrm{B}}$
is
\begin{align}
  &\hat{h}_{\mathrm{SP}} = \left[
  -\frac{ \hbar^2 }{2} \sum_{\nu}
    \frac{1}{m_\nu} \frac{\partial^2}{\partial \nu^2} 
  - \mu_F \right] \hat{\sigma}_0, \\[1mm]
  &\hat{h}_{\mathrm{SO}} 
	= V_{\mathrm{SO}} \delta (z) \bs{e}_z \cdot 
	  \left[ \bs{p} \times \hat{{\bs{\sigma}}} \right], \hspace{4mm}
  \hat{h}_{\mathrm{B}} 
	= V_{\mathrm{B}} \delta (z) \hat{{\sigma}}_0, 
\end{align}
where $\hat{h}_{\mathrm{SO}}$ and $\hat{h}_{\mathrm{B}} $ represent
the Rashba spin-orbit interaction (RSOI) and the potential barrier at the
interface and $m_\nu$ are the effective mass in the $\nu$ direction.
The Fermi surface of the SC and the N is uniaxially anisotropic and
isotropic respectively. Introducing the anisotropic effective mass,
one can model anisotropic FSs. Their effective masses are given by 
\begin{align}
  (m_a, m_b, m_c) = 
	\left\{ \begin{array}{cl}
	  (m_N, m_N, m_N) & \text{~for~} x <    0, \\[1mm]
	  (m_\parallel, m_\parallel, m_\perp)  & \text{~for~} x \geq 0. \\
	\end{array} \right.
\end{align}
These assumptions are valid for layered superconducting materials such
as Sr$_2$RuO$_4$. 

In this paper, we consider four types of pair potentials: 
(1) spin-singlet $d_{zx}+id_{yz}$-wave, 
(2) spin-triplet $f$-wave, 
(3) spin-triplet helical $p$-wave, and 
(4) spin-singlet $s$-wave. 
Each pair potential is given by 
\begin{align}
  & \text{(1)}~d_0    = \bar{\Delta}_0 ( \partial_c \partial_a + i \partial_b \partial_c) / k^S_{\parallel}k^S_{\perp}, \\[0mm]
  & \text{(2)}~\bs{d} = \bs{e}_a (\bar{\Delta}_0\partial_a \partial_b \partial_c )     /(k^S_{\parallel})^2k^S_{\perp},\label{eq:f-wave}\\
  & \text{(3)}~\bs{d} = \bar{\Delta}_0 ( \bs{e}_a \partial_a + \bs{e}_b \partial_b)     / k^S_{\parallel}, \\[0mm]
  & \text{(4)}~d_0    = \bar{\Delta}_0, 
\end{align}
where $\bar{\Delta}_0$ is determined so that
$\mathrm{max}[d_{\bs{k}}]=\Delta_0$ on the FS with 
$d_{\bs{k}} = \sqrt{d_0^2 + |\bs{d}|^2}$ and 
$\Delta_0 \in \mathbb{R}$ characterizes the amplitude of the
pair potential, and
$k_{s\parallel}$ and $k_{s\perp}$ are the Fermi
momentum parallel and perpendicular to the $k_a$-$k_b$ plane in the
superconductor.  
The $d$-vector of the spin-nematic $f$-wave pairing is 
assumed parallel to the $a$-axis. This $d$-vector
reproduces the NMR results \cite{Pustogow_arXiv_2019,
Ishida_arXiv_2019}, where an external magnetic field is applied in
the [100] direction. This anisotropic $d$-vector 
reduces the four-fold rotational
symmetry stemming from the crystal structure into two-fold one (i.e., spin-nematic
superconductivity). The spin-nematic $f$-wave pairings with $\bs{d} \parallel \bs{a}$ and  
$\bs{d} \parallel \bs{b}$ are degenerated. 
In this paper, we refer to each pair potential as 
(1) chiral $d$-wave (chiral DW), 
(2) spin-nematic $f$-wave (spin-nematic FW), 
(3) helical $p$-wave (helical PW), and 
(4) $s$-wave (SW) SC. 
The SW, helical PW, and spin-nematic FW pairings are time-reversal
symmetric, whereas the
the chiral DW one breaks TRS. 

From the experimental data obtained so far, we cannot exclude the
possibility of the existence of subdominant pair potentials for the FW
pairing where the subdominant component should also belong to the 2D
irreducible representation\footnote{In this paper, we consider
subdominant component only for the FW pairing. Subdominant components
in a spin-triplet superconductor can change the direction of the
$d$-vector, resulting in the mixing between spin subspace. Such spin
mixing can affects the surface ABSs.}. Such a subdominant component changes the
node to a small minimum and can gain the condensation energy.
Therefore, to discuss the effects of a subdominant component, we
introduce the parameters $\eta_a$ and $\eta_b$ with which the
$d$-vector is given by  
\begin{align}
\bs{d} = 
\bar {\Delta}_0     \bs{e}_a (k_a k_b k_c) 
+ \eta_a {\Delta}_0 \bs{e}_c k_a
+ \eta_b {\Delta}_0 \bs{e}_c k_b, 
\label{eq:subdomi}
\end{align}
where, we consider the subdominant components linear function of the
momentum (i.e., $p_x$- and $p_y$-wave
pairing as subdominant components with $\bs{d} \parallel \bs{c}$). 


\begin{figure}[tb]
	\centering
  \includegraphics[width=0.48\textwidth]{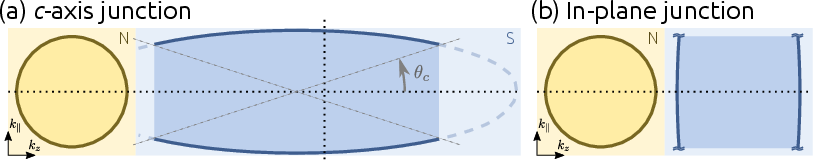}
	\caption{Schematics of the Fermi surfaces.  
	The FS of the SC is anisotropic due to the anisotropic effective
	mass. The cut-off angle $\theta_c$ is introduced to make the FS in
	the SC cylindrical which is important when we estimate the effects
	of the Rashba spin-orbit coupling.}
	\label{fig:Sche_FS}
\end{figure}

The wave functions obey the Bogoliubov-de Gennes (BdG) equation: $\check{H} \Psi = E \Psi$.
In the present case, the momenta parallel to the interface $k_x$ and
$k_y$ are good quantum numbers because of translational symmetry.
Therefore, the BdG equation is decomposed as 
\begin{align}
  & \check{H}_{\bs{k}_\parallel} \Psi_{\bs{k}_\parallel}(z)
	= E_{\bs{k}_\parallel} \Psi_{\bs{k}_\parallel}(z) 
	\\[0mm]
	&\check{H}_{\bs{k}_\parallel}(z) 
	=
	\left[
		\begin{array}{cc}
			 \hat{h     } _{ \bs{k}_\parallel}(z)    ~ &
			 \hat{\Delta} _{ \bs{k}_\parallel}(z)\\[2mm]
			-\hat{\Delta}^*_{-\bs{k}_\parallel}(z)&
			-\hat{h     }^*_{-\bs{k}_\parallel}(z)    ~ \\
	\end{array}\right], 
	\\[0mm]
  &\Psi(\bs{r})
  =
  \sum_{\bs{k}_\parallel}
  \Psi_{\bs{k}_\parallel}(z)
  \frac{  e^{i (k_x x + k_y y) } }{\sqrt{L_x L_y}}, \\
  &\Psi_{\bs{k}_\parallel}(z)
  =
	[~
   \psi_{\ua, { \bs{k}_\parallel}}        ~ ~
	 \psi_{\da, { \bs{k}_\parallel}}        ~ ~
	 \psi_{\ua, {-\bs{k}_\parallel}}^\dagger~ ~
   \psi_{\da, {-\bs{k}_\parallel}}^\dagger~ 
  ]^T, 
\end{align}
where $\bs{k}_\parallel = (k_x, k_y, 0)$.  The normal part of 
$\check{H}_{\bs{k}_\parallel}$ is given by
\begin{align}
	&\hat{h}_{ \bs{k}_{\parallel}} = 
  \left[ -\frac{ \hbar^2 }{2 m_z} \partial z^2
  - \mu_{\bs{k}_\parallel}  \right] \hat{\sigma}_0 
	+\hat{V} \delta (z)
	\\[0mm]
	& \mu_{\bs{k}_\parallel} 
	= \mu_F 
	+ { \hbar^2 k_x^2 }/{2 m_x}
	+ { \hbar^2 k_y^2 }/{2 m_y}, \\[1mm]
	& \hat{V} = V_{\mathrm{B}} \check{\sigma}_0
	+ V_{\mathrm{SO}} \hat{\Lambda}, 
\end{align}
where $\hat{\Lambda} = \bs{e}_z \cdot \left[ \bs{p} \times
\hat{{\bs{\sigma}}} \right]$. 
In what follows, we make $\bs{k}_\parallel$ explicit only when
necessary. 

To obtain the wave functions in the junction, we first solve the BdG
equation in each region. When a quasiparticle with the spin $\alpha =
$ $\ua$ or $\da$ is injected into the interface, 
the wave function in the N region can be written as a linear
combination of every possible wave functions: 
\begin{align}
  &\Psi^N(z)
  = e^{+i k^N_z \check{\tau}_3 z} \vec{a}_\alpha
	+ e^{-i k^N_z \check{\tau}_3 z} \vec{r}_\alpha
	\\[0mm]
	& \vec{a}_\alpha =
	\left\{ \begin{array}{ll}
	( ~ 1~ ~0 ~ ~0 ~ ~0 ~)^T & \text{~ ~ for ~}\alpha =\, \ua \\
	( ~ 0~ ~1 ~ ~0 ~ ~0 ~)^T & \text{~ ~ for ~}\alpha =\, \da \\
	\end{array} \right. 
	\\
	& \vec{r}_\alpha = ( 
    ~r^{p}_{\ua \alpha}~
    ~r^{p}_{\da \alpha}~
    ~r^{h}_{\ua \alpha}~
    ~r^{h}_{\da \alpha}~
  )^T, 
\end{align}                  
where $r^{p(h)}_{\alpha' \alpha}$ is the normal (Andreev) reflection
coefficients. The momentum in the $z$-direction is given by $k^N_z =
\sqrt{2 m_N \mu_{\bs{k}_\parallel}}/\hbar$ where we have used the
Andreev approximation valid when $\mu \gg \Delta_0$, which allows us
to ignore the energy dependence of the momentum. 
The wave function in the SC is given by 
\begin{align}
  & \Psi^S(z)
	= \check{U} e^{i k^S_z \hat{\tau}_3 z} \vec{t}, \\[0mm]
	& \check{U} = 
	\left[ \begin{array}{cc}
		u_0 \hat{\sigma}_0            & v_0 \hat{\Delta}_o/d_{\bs{k}}  \\[1mm]
		v_0 \hat{\Delta}_o^\dagger/d_{\bs{k}} & u_0 \hat{\sigma}_0          \\[1mm]
  \end{array} \right] \\[0mm]
	& \vec{t}_\alpha = (
	~ t_{1\alpha}^p,~ ~ t_{2\alpha}^p,~ 
  ~ t_{1\alpha}^h,~ ~ t_{2\alpha}^h~)^T, 
\end{align}
where we have used the Andreev approximation; 
$k^S_z = \sqrt{2m_z \mu_{k_\parallel}}/\hbar$. The
$t_{1(2)}^{p(h)}$ coefficients are the transmission coefficients where
the superscript indicates the transmission as a particle-like or
hole-like quasiparticle and the subscript does the band index.

The differential conductance can be obtained from the reflection coefficients
as in Blonder-Tinkham-Klapwijk (BTK) theory \cite{BTK}. To obtain the
reflection coefficients, we need to
match the wave functions at the interface $z=0$. There are two
boundary conditions to conserve the
probability density: 
\begin{gather}
    \Psi^N(z)\Big|_{z=0} 
	= \Psi^S(z)\Big|_{z=0}, 
	\\[1mm]
    \lim_{\gamma \rightarrow 0} \int_{-\gamma}^{\gamma} \check{H} \Psi(z) dz
  = \lim_{\gamma \rightarrow 0} \int_{-\gamma}^{\gamma}        E  \Psi(z) dz. 
\end{gather}
Substituting the wave function in each region, we obtain the first
boundary condition in terms of the coefficients
\begin{align}
  \vec{a}_\alpha + \vec{r}_\alpha
  = \check{U} \vec{t}_\alpha, 
  \label{eq:bc01}
\end{align}
and the second boundary condition
\begin{align}
  & ( \vec{a}_\alpha - \vec{r}_\alpha )
	- \check{V}_S \check{U} \vec{t}
	= 0, 
  \label{eq:bc02}
	\\
	& \check{V}_S = 
	\bar{v}_S \check{U}' \check{U}^{-1}
	+2i \check{\tau}_3 \check{V} / \hbar v_N , 
\end{align}
where $\check{U}' = \check{\tau}_3 \check{U}\check{\tau}_3$, 
$v_{N(S)} = \hbar k^{N(S)}_z / m_z$ are the velocities 
in the N (S), and $\bar{v}_{S} = v_S / v_N$. 
Combining the equations
\eqref{eq:bc01} and \eqref{eq:bc02}, we obtain the following equation 
\begin{align}
  \vec{r}_\alpha 
	&= 
  \left( \check{\tau}_0 +\check{V}_S \right)^{-1} 
  \left( \check{\tau}_0 -\check{V}_S \right)
	\vec{a}_\alpha. 
\end{align}
Calculating $\vec{r}_\alpha$ numerically, we can obtain the reflection
coefficients for each reflection process. 

The differential conductance $G(eV) = dI/dV$ can be calculated by 
BTK formula \cite{BTK}. The conductance is give by  
\begin{align}
  & G (eV) = \sum_{\bs{k}_\parallel,\alpha}
	g_{\alpha}(E=eV,\bs{k}_\parallel)
	\Theta(\theta_s - \theta_c)
	\\
	& g_{\alpha}
	= v_N [ 1 - (\vec{r}_\alpha)^\dagger \check{\tau}_3 \vec{r}_\alpha], 
\label{eq:BTK}
\end{align}
where $g_{\alpha}$ is the partial conductance. In the cylindrical-FS model, we
introduce the cut-off angle $\theta_c$ as shown in
Fig.~\ref{fig:Sche_FS}(a). When $\theta_s < \theta_c$ with $\tan \theta_s
= k_\parallel/k^S_z$, the partial charge current cannot flow the
junction.  Throughout this paper,
we consider the zero-temperature.  

\section{Tunnelling spectroscopy along $c$ axis}

\begin{figure}[tb]
	\centering
  \includegraphics[width=0.48\textwidth]{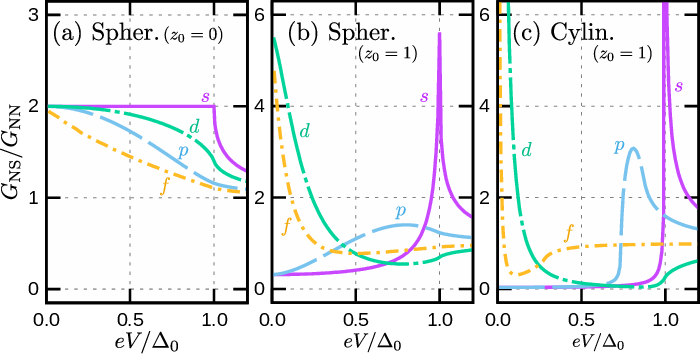}
	\caption{Conductance of the NS junction along the $c$-axis.  The
	indices $s$, $p$, $d$, and $f$ means the SW, helical PW, chiral DW, and FW
	pairing respectively.  The Fermi surface is spherical in (a) and
	(b), and cylindrical in (c). The normalized effective masses in (c)
	are set to $(\bar{m}_x, \bar{m}_y,\bar{m}_z) = (1.3, 1.3, 16.0)$.
	The barrier potential is set to $z_0 = 0$ in (a) and $z_0 = 1$ in
	(b) and (c). The conductance is normalized to its value in the
	normal state $G_{\mathrm{NN}}$. The result for chiral PW is the identical
	to that for helical PW in the absence of spin-dependent potentials. }
	\label{fig:Gns_caxis}
\end{figure}

The differential conductance along the $c$-axis is shown in
Fig.~\ref{fig:Gns_caxis}. 
The conductance $G_{\mathrm{NS}}$ is normalized to its value in the normal state $\gnn$,
which is obtained by setting $\Delta_0 = 0$.  The FS is
spherical in Figs.~\ref{fig:Gns_caxis}(a) and \ref{fig:Gns_caxis}(b)
and cylindrical in Fig.~\ref{fig:Gns_caxis}(c), where the effective
mass\cite{OldeOlthof_PRB_2018} is respectively set to
$(\bar{m}_x,\bar{m}_y,\bar{m}_z) = (1.0,1.0,1.0)$ and $(1.3,1.3,16.0)$ with
$\bar{m}_\nu = m_\nu/m_N$. The barrier potential is $z_0=0$ in
Fig.~\ref{fig:Gns_caxis}(a) and $z_0 = 1.0$ in
Figs.~\ref{fig:Gns_caxis}(b) and \ref{fig:Gns_caxis}(c). 

When the FS is spherical, $G_{\mathrm{NS}}$ without the
barrier is larger than $G_{\mathrm{NN}}$ within the gap as shown in 
Fig.~\ref{fig:Gns_caxis}(a). In this case, the normal reflection is
forbidden because there is no barrier and no Fermi-momentum mismatch. 
As a result, the injected quasiparticle within the gap propagates 
into the SC as a Cooper pair 
with the charge $2e$. Therefore, the conductance $\gns$ must be larger than
$\gnn$. In an unconventional SC, the gap size depends on $\bs{k}$
and can be smaller than $\Delta_0$, which changes the
conductance spectra depending on the node type. 
The helical PW has the point nodes at $k_a=k_b=0$ and show the dome-shape
$\gns$ \cite{OldeOlthof_PRB_2018}. 
The chiral DW has a line node at $k_c=0$ in addition to the point
node. However, its gap amplitude $d_{\bs{k}}$ maximizes at $k_c/
k^s_{\perp} =
1/\sqrt{2}$ which results in a larger $\gns$ than that of the helical
PW case where $d_{\bs{k}}$ maximizes at $k_c =0$ (i.e., at the
velocity $v_N=0$).  The FW has line nodes at $k_a = 0$ and $k_b = 0$ which
results in the sharper ZEP. 

The barrier potential changes the conductance spectra drastically as
shown in Fig.~\ref{fig:Gns_caxis}(b). The conductance for the chiral DW and
FW junctions have sharp ZEPs due to the resonant tunneling through the
zero-energy ABSs at the interface 
similar to in-plane tunneling of $d_{xy}$-wave \cite{TK95} and 
$p_{x}$-wave junctions \cite{YTK98,Yakovenko}. The pair
potential of the chiral DW and FW SCs is antisymmetric under $k_c
\leftrightarrow -k_c$ which results in the ABSs at the (001) surface
\footnote{The ABSs are formed by the interference between the
quasiparticles propagating with $k_z$ and $-k_z$. The ABSs are present
when the phases of the pair potentials $\Delta(\bs{k}_\parallel, k_z)$
and $\Delta(\bs{k}_\parallel,- k_z)$ are different. In particular, the
ABS appears at the zero energy when the condition
$\phi(\bs{k}_\parallel, k_z) -\phi(\bs{k}_\parallel, k_z)= \pi$ is
satisfied, where $\phi$ is the phase of the pair potential 
\cite{Hu,TK95,ABSR1,ABSR2}}. 
In contrast, the conductance for the helical PW is V-shaped
at the low energy reflecting the point node. The coherence
peak around $|eV| \sim 0.7 \Delta_0$ is broad because of the
angle-dependent pair potential. The conductance spectra for
the SW is well-known U-shaped one in the presence of the barrier
potential. 

When the FS in the SC is cylindrical, 
the node structure near $k_a = k_b = 0$ cannot contribute to the
transport. Moreover, the channels relevant to the transport
is restricted [see Fig.~\ref{fig:Sche_FS}(a)]. Consequently, as indicated in 
Fig.~\ref{fig:Gns_caxis}(c), $\gns$ for the helical PW
changes from V-shape to U-shape because the point node does not
contribute to the transport. The characteristic energy scale for
the FW is changed: a kink appears around $|eV| \sim 0.4 \Delta_0$. 
In the FW junction, 
the channels with $d_{\bs{k}}=\Delta_0$ cannot contribute to the transport due to
the cutoff for modelling the cylindrical FS of
Sr$_2$RuO$_4$\footnote{The energy scale depends also on the size of the FS in the N region}.
Therefore, the conductance structure appears only
at $|eV| < 0.4 \Delta_0$. The ZEPs for the chiral DW and FW becomes
narrower than those for the spherical-FS case because the
Fermi-momentum mismatch reduces the transparency at the interface.

\subsection{Robust zero-energy peak of chiral $d$-wave junction}
\begin{figure}[tb]
	\centering
  \includegraphics[width=0.48\textwidth]{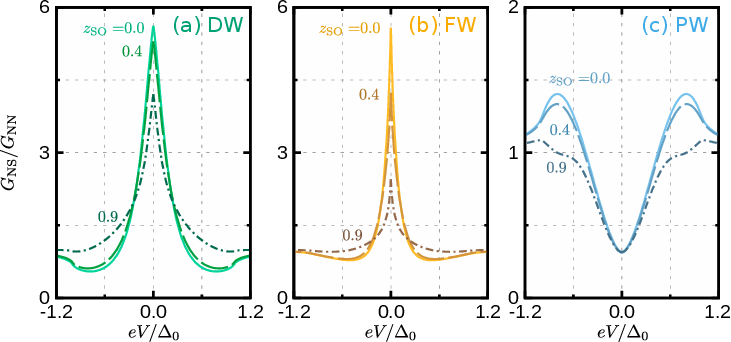}
	\caption{Effects of Rashba spin-orbit interaction on the conductance 
	along $c$-axis with \textit{spherical} Fermi
	surface.  The strength of the RSOI are set to $z_{\mathrm{SO}} = 0$, $0.4$, and $0.9$. The barrier potential is set to $z_0 = 1$. The
	conductance is normalized to its value in the normal state
	$G_{\mathrm{NN}}$.}
	\label{fig:Gns_SOC_Sphe}
\end{figure}

Near an N/SC interface, the parity mixing occurs due to the RSOI. 
To discuss the robustness of the zero-energy peak against
parity mixing,
we calculate $\gns$ taking into account the 
RSOI at the interface \cite{Samokhin, OldeOlthof_PRB_2018}. 
The RSOI at an interface changes $\gns$ significantly. For
instance, it is demonstrated that the ZEP of
spin-singlet chiral SCs are robust against the RSOI, whereas those of
spin-triplet chiral SCs can be suppressed \cite{Kobayashi_PRB_2015}.
The effects of the RSOI on the conductance with the \textit{spherical}
FS are shown in Fig.~\ref{fig:Gns_SOC_Sphe}, where $z_0 = 1$ and the
pairing is assumed (a) FW, (b) chiral DW, and (c) helical PW pairings.  The ZEP
of the spin-singlet SC can survive even in the presence of the strong
RSOI as shown in Fig.~\ref{fig:Gns_SOC_Sphe}(a). In the present case,
the ZEP of the
triplet SC can survive even in the presence of the strong RSOI as
shown in Fig.~\ref{fig:Gns_SOC_Sphe}(b) because we consider a
different situation from Ref.~\onlinecite{Kobayashi_PRB_2015}. In the
present case, the $d$-vector is $ \bs{d} \perp \bs{z}$, whereas $
\bs{d} \parallel \bs{z}$ in their case. The coherence peak for the
helical PW is suppressed by the ROSI as shown in
Fig.~\ref{fig:Gns_SOC_Sphe}(c). 
\begin{figure}[tb]
	\centering
  \includegraphics[width=0.48\textwidth]{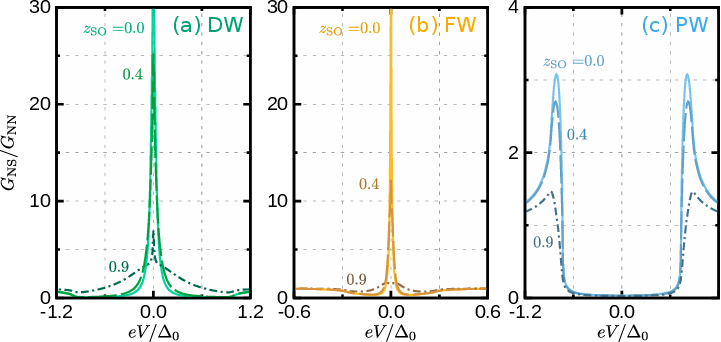}
	\caption{Effects of Rashba spin-orbit interaction on the conductance with
	\textit{cylindrical} Fermi surface. The results are plotted in the
	same manner as in Fig.~\ref{fig:Gns_SOC_Sphe}. The parameters are
	set to the same values as the corresponding panels in
	Fig.~\ref{fig:Gns_SOC_Sphe}. }
	\label{fig:Gns_SOC_Cylin}
\end{figure}

When the FS is cylindrical, the channels with small
$|\bs{k}_\parallel|$
cannot contribute to the transport [see Fig.~\ref{fig:Sche_FS}(a)]. In
other words, the charge current is mainly carried by the channels with
the stronger RSOI whose amplitude is proportional to
$|\bs{k}_\parallel|$. The conductance with the cylindrical FS is shown in
Fig.~\ref{fig:Gns_SOC_Cylin}. The ZEP for the spin-singlet chiral DW
is robust against the RSOI, whereas that for the FW is fragile. Their
peak heights at $z_{\mathrm{SO}} = 0.9$ are $\gns(eV=0) \sim 7 \gnn$
and $1.8\gnn$, respectively. The conductance of the helical PW
junction does not qualitatively depend on the shape of the FS. 

\begin{figure}[tb]
	\centering
  \includegraphics[width=0.40\textwidth]{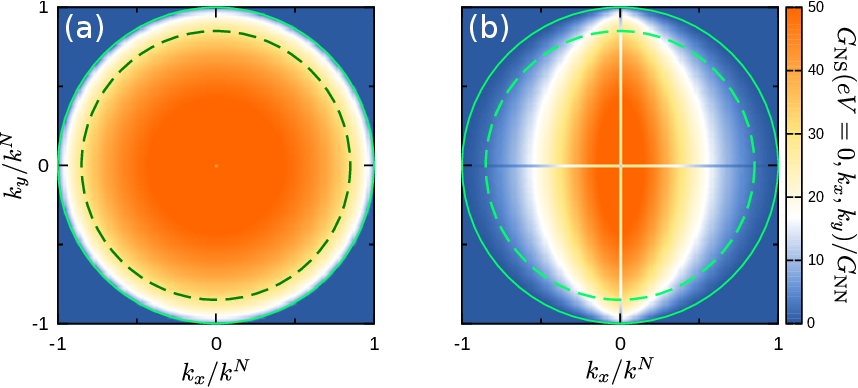}
	\caption{Effects of Rashba spin-orbit interaction on 
	angle-resolved zero-energy conductance of $c$-axis
	junction. The 
	chiral DW and spin-nematic FW pairings are used in (a) and (b), respectively.  
	The circles with the solid and broken line indicate the Fermi
	surface of the N and the smallest radius of the cylindrical Fermi
	surface in S. In the cylindrical-FS model, only the channels
	between solid and broken circles can contribute the conductance. }
	\label{fig:ARG}
\end{figure}

The angle-resolved zero-energy $\gns$ for the chiral DW
and spin-nematic FW are respectively 
shown in Figs.~\ref{fig:ARG}(a) and \ref{fig:ARG}(b), where $(\bar{m}_x,\bar{m}_y,\bar{m}_z) =
(1.3,1.3,16.0)$, $z_0 = 3.0$, and $z_{\mathrm{SO}} = 1.0$.  The circles with the solid and broken line in
Fig.~\ref{fig:ARG} indicate the maximum $|\bs{k}_\parallel|$ in the N and the
minimum $\bs{k}_\parallel$ in the SC due to the cutoff. In the
cylindrical-FS model, only the channels between the solid and broken
circles can contribute to $\gns$. 
The zero-energy ABS for the chiral DW state is robust against the
RSOI, while that for the FW one is substantially suppressed. This
difference comes from the difference in parity of these pairings: the
chiral DW (spin-nematic FS) pairing is even (odd) under inversion. For
an even-parity SC, a line node can be topologically stable and
correspondingly a robust ABS with a flat band arises at an
interface.\cite{Kobayashi_PRB_2015} On the other hand, no
topologically stable line node exists for an odd-parity SC with a
spin-orbit interaction, and therefore a zero-energy ABS is fragile
against the RSOI. 
\begin{figure}[tb]
	\centering
  \includegraphics[width=0.40\textwidth]{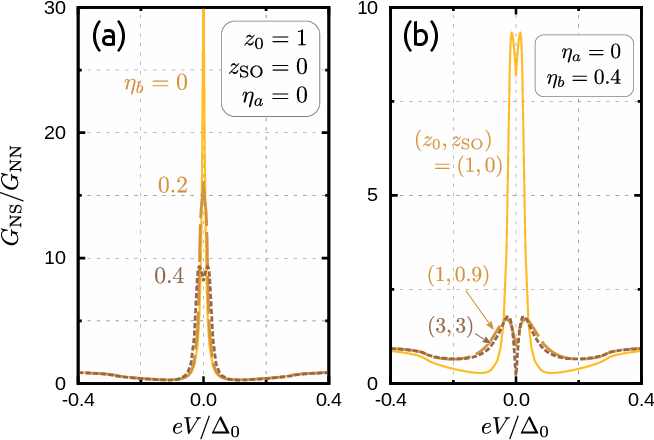}
	\caption{Effects of the subdominant component $\bs{z}p_y$ on
	$\gns$ of the FW junction.  A subdominant component is added to the
	FW pairing as given in Eq.~\eqref{eq:subdomi}. In the calculations,
	we set $\eta_a=0$ and $\eta_b \neq 0$ reflecting the nematic
	superconducting state. The results for $\eta_a \neq 0$ and $\eta_b = 0$
	are not qualitatively different from those shown in the figures
	above. The cylindrical FS is used. 
	(a) The subdominant component splits 
	the ZEP, where the parameters are set to $z_0=1$ and $z_{\mathrm{SO}}=0$. 
	(b) When both of the subdominant component and the RSOI exist,
	$\gns$ can be even a narrow V-shaped one, where the parameters are
	set to $(z_0,
z_{\mathrm{SO}})=(1,0)$, $(1,0.9)$, or $(3,3)$
	}
	\label{fig:Gns_subdomi}
\end{figure}

The fragility of the ZEP for the FW pairing is more prominent when
taking into account the subdominant pairing state in
Eq.~\eqref{eq:subdomi}.  The effects of the subdominant component on
$\gns$ of the FW junction are shown in Fig.~\ref{fig:Gns_subdomi},
where the cylindrical FS is employed.  In the calculations, we have
used asymmetric parameter (i.e., $\eta_a=0$ and $\eta_b \neq 0$)
reflecting the nematic nature. The results with $\eta_a \neq 0$ and
$\eta_b = 0$ are not qualitatively different from those shown in this
paper. As shown in Fig.~\ref{fig:Gns_subdomi}(a), the ZEP is split and
suppressed by the subdominant component even when $z_{\mathrm{SO}}=0$. 
When both of the subdominant component and the RSOI exist, the
low-energy spectrum can be V-shaped even though the width of the
structure is narrower than $\Delta_0$ as shown in
Fig.~\ref{fig:Gns_subdomi}(b), where $\eta_b = 0.4$ and $(z_0,
z_{\mathrm{SO}})=(1,0)$, $(1,0.9)$, or $(3,3)$. 

So far, ABSs have never been experimentally observed in the (001)
surface of Sr$_2$RuO$_4$ \cite{Suderow_2009, Firmo_PRB_2013, Madhavan}. 
Therefore, we conclude that the $d_{zx} + i d_{yz}$-wave does not 
explain the transport measurements along the $c$-axis of
Sr$_2$RuO$_4$.  The spin-nematic FW and helical PW pairings remain as possible
pairing symmetry of Sr$_2$RuO$_4$. 

\section{In-plane tunneling spectroscopy}
\begin{figure}[tb]
	\centering
  \includegraphics[width=0.48\textwidth]{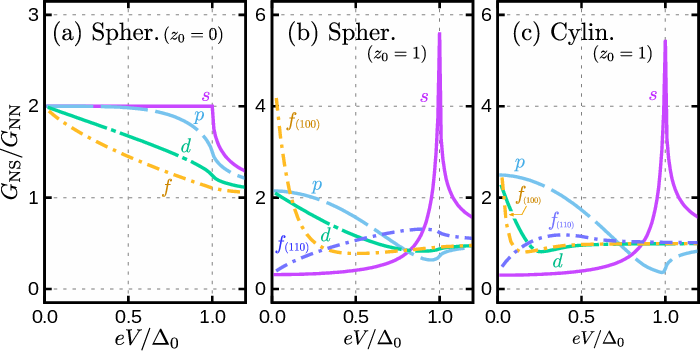}
	\caption{Conductance of the NS junction with $\bs{c} \perp \bs{z}$. 
	The results are plotted in the same manner as in
	Fig.~\ref{fig:Gns_caxis}. In the in-plane junction, $\gns$ for the
	FW SC depends on the direction of the interface; ZEP for (100) and
	V-shaped for (110).  }
	\label{fig:Gns_aaxis}
\end{figure}

The in-plane tunneling spectroscopy can distinguish the spin-nematic
FW and helical PW without ambiguity. The results 
are shown in Fig.~\ref{fig:Gns_aaxis}. The parameters are set to 
(a) $(z_0, \bar{m}_x, \bar{m}_y, \bar{m}_z) = (0, 1, 1, 1) $, 
(b) $                                         (1, 1, 1, 1) $, and 
(c) $                                         (1, 1.3, 16, 1.3) $. 
In the absence of $z_0$, the conductance $\gns$ for the helical PW, chiral DW, and FW are
dome-shaped ZEP, ZEP, 
and ZEP, respectively.
When $z_0 \neq 0$, $\gns$ shows different behavior depending on the
pairing symmetry. In particular, $\gns$ for the FW significantly
depends on the direction of the junction. As shown in
Fig.~\ref{fig:Gns_aaxis}(b), $\gns$ for the FW are the ZEP or V-shaped
dip in the (100) and (110) junctions, respectively. In the helical PW and
chiral DW junctions, $\gns$ do not depend on the 
direction of the junction
\footnote{In the square-lattice tight-binding model,
$\gns$ depends on the direction reflecting the FS anisotropy in the
$ab$ plane} as in the chiral $p$-wave junction\cite{Honerkamp98}: both of them show the broad ZEP. 
Comparing Figs.~\ref{fig:Gns_aaxis}(b) and \ref{fig:Gns_aaxis}(c), we
see that the shape of the FS in the SC does not change $\gns$
qualitatively in the in-plane junction. 
Note that, when the FS is cylindrical, the characteristic energy scale
for chiral DW
and FW are smaller than those for helical PW and SW because the
channels with $d_{\bs{k}}=\Delta_0$ cannot contribute the
transport in the chiral DW and FW cases [see Fig.~\ref{fig:Sche_FS}(b)]. 
\section{Spin-nematic $f$-wave versus non-chiral $d_{xy}$-wave}

\begin{figure}[tb]
	\centering
  \includegraphics[width=0.47\textwidth]{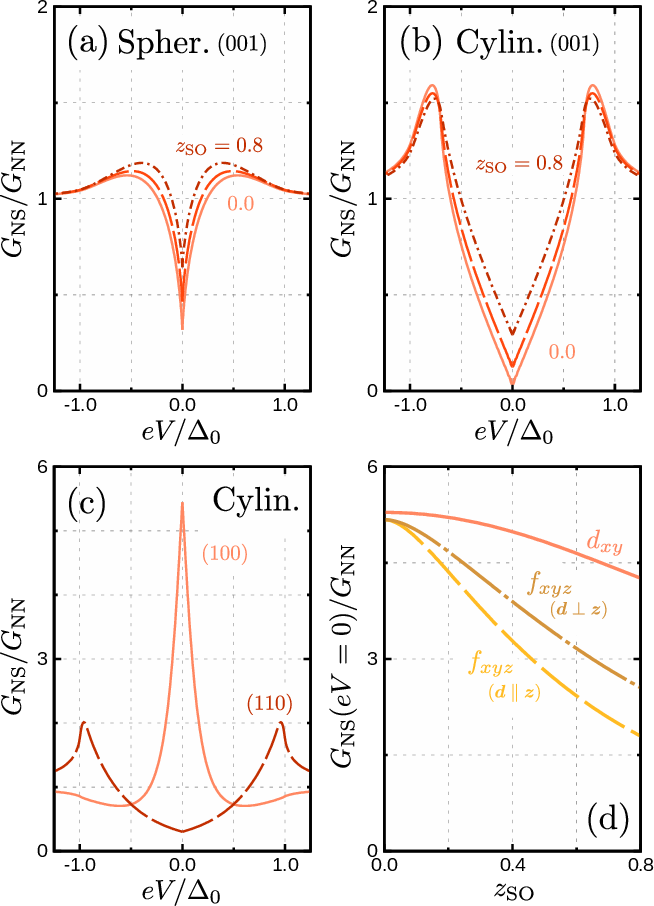}
	\caption{(a)(b) Conductance of the \textit{non-chiral} $d$-wave junction
	with $\bs{c} \parallel \bs{z}$. The width of the V-shaped is
	characterized by $\Delta_0$. (c) Conductance of the \textit{non-chiral} $d$-wave junction 
	with $\bs{c} \perp \bs{z}$. 
	(d) $z_{\mathrm{SO}}$-dependences of
	$\gns|_{eV=0}$ for $d_{xy}-$ and $f_{xyz}$-wave pairings. The
	zero-energy peak for $d_{xy}$-wave junction is more robust against spin mixing
	than those for $f$-wave. 
	The barrier parameters are set to
	$z_0=1$ in (a), (b), (c), and (d). The spin-orbit interaction is set
	to $z_{\mathrm{SO}}=0.0$, $0.6$, and $0.8$ in (a) and (b), 
	and $z_{\mathrm{SO}}=0.0$ in (c). 
	}
	\label{fig:Gns_dw}
\end{figure}

\begin{table*}[tb]
  \centering
	\caption{Summary of conductance spectra. The spectra 
	strongly depends on the paring symmetry and the direction of the
	junction. The conductance spectra are classified into three types:
	zero-energy peak (ZEP), dome-like broad peak (Dome), U-shaped dip
	(U), and V-shaped dip (V). 
	The helical PW is consistent with the NMR measurements only
	qualitatively but not quantitatively. }
	\vspace{2mm}
  \includegraphics[width=0.78\textwidth]{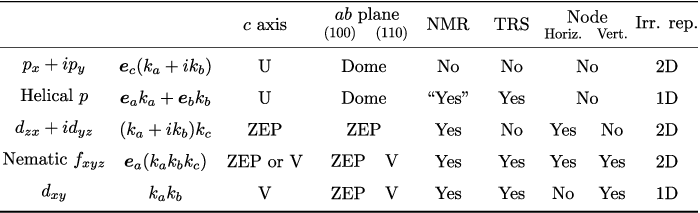}
	\label{Tab}
\end{table*}
The comparison between the spin-nematic $f_{xyz}$-wave and
spin-singlet \textit{non-chiral} $d_{xy}$-wave pairings would provide useful
information to discuss the pairing symmetry of Sr$_2$RuO$_4$ even
though the non-chiral $d_{xy}$-wave pairing did not explain
several experiments. 
The conductance spectra
$\gns$ of the $d_{xy}$-wave junction are known to be similar to those
of the FW junction: $\gns$ shows the V-shaped, ZEP, and
V-shaped spectra in the (001)-, (100)-, and (110)-interface junctions. 

The conductance spectra along the $c$-axis of the \textit{non-chiral} $d$-wave
junctions are shown in Fig.~\ref{fig:Gns_dw}(a) and
\ref{fig:Gns_dw}(b), where the spherical and cylindrical FS is used in
Fig.~\ref{fig:Gns_dw}(a) and \ref{fig:Gns_dw}(b) respectively and the
$d_{xy}$-wave pair potential is given by $ d_0    = \bar{\Delta}_0 (
\partial_a \partial_b ) / (k^S_{\parallel})^2 $.  In both cases,
$\gns$ is V-shaped regardless of $z_{\mathrm{SO}}$. 
Differing from the FW case in
Fig.~\ref{fig:Gns_subdomi}(b), the width of the V-shaped structure is
about $\Delta_0$ in the $d_{xy}$-wave case.
The V-shaped
structure is more prominent in the cylindrical-FS model [i.e.,
Fig.~\ref{fig:Gns_dw}(b)] because channels near the intersection
of the two line nodes ($k_x=k_y=0$) do not contribute to the
transport because of the cutoff $\theta_c$.  

The conductance spectra of in-plane junctions (i.e., $\bs{c} \perp
\bs{z}$) are shown 
in Fig.~\ref{fig:Gns_dw}(c). The transport in this case have been 
established\cite{Tanaka_PRL_1995}: $\gns$ becomes ZEP and V-shaped in
the (100)- and (110)-interface junctions respectively. The
$z_{\mathrm{SO}}$-dependences of $\gns|_{eV=0}$ for the $d_{xy}$- and
spin-nematic $f_{xyz}$-wave pairings are shown in 
Fig.~\ref{fig:Gns_dw}(d), where $d$-vector for the FW is assumed 
$\bs{d} \parallel \bs{a}$ or $\bs{d} \parallel \bs{b}$. The ZEP for the
$d_{xy}$ is more robust against the spin-mixing than those of $f_{xyz}$-wave junctions as
shown in Fig.~\ref{fig:Gns_dw}(d).

\section{Discussions}

The conductance spectra for each pairing and each junction direction
are summarized in Table~\ref{Tab}, where the 
conductance spectra are classified into the four types: 
zero-energy peak 
(ZEP), dome-like broad peak (Dome), U-shaped dip (U), and V-shaped dip
(V). The tunneling spectroscopy of the (001)-, (100)-, and
(110)-interface junctions is found to be important clues to identify the pairing
symmetry of Sr$_2$RuO$_4$. 
In particular, it should be emphasized that the $c$-axis transport
measurements by the STM\cite{Suderow_2009, Firmo_PRB_2013, Madhavan} are inconsistent with the
chiral DW scenario, but support both of the spin-nematic FW and helical PW. 
These pairings can clearly be
distinguished by transport measurements of the (100) and (110) junctions. 
Such a direction-dependent $\gns$ has been well
established for high-$T_c$ $d$-wave SCs. In the high-$T_{c}$ SC, the
ZEP appears in the (110) junction and does not in the (100) junction
\cite{TK95, Experiment4, Experiment5}. 

We did not take multiband effects into account. However, the
multiband effect does not change $\gns$ qualitatively but
quantitatively\cite{Yada_JPSJ_2014}. In particular, for the (001)
junction,  the interaction among the bands would not play a substantial role in
transport in the [001] direction since the energy bands are less
dispersive with respect to $k_c$ and do not overlap each other when
they are projected in the $k_a$-$k_b$ plane.
The conductance spectra in in-plane
junctions, would be modified by the multiband effects more
significantly compared with those of the (001) junction. However, the
conductance spectra in the [100] and [110] directions for the helical
PW (spin-nematic FW) pairing are expected to be \textit{qualitatively}
identical (different) because the multiband effects change the
spectra only qualitatively.  It would be interesting to calculate
$\gns$ of the chiral $d$-wave, helical $p$-wave, and spin-nematic
$f_{xyz}$-wave junctions with taking the multiband effect into
account. 

The spatial dependence of the pair potential, which is caused by the
surface reconstruction, the interface reflection\cite{Barash97,
Matsumoto_JPSJ_1999, Furusaki_PRB_2001, Nagato_JLTP_1993}, and
interface roughness\cite{Nagato_JLTP_1998, SIS3, Bakurskiy_2017,
Tanuma01}, is not taken into account in the present calculations. The
spatial dependence would change $\gns$ quantitatively but not
qualitatively.  Therefore, our conclusion would be valid even if the
pair potential is spatial dependent.

\section{Summary}
In this paper, we have proposed 
that the tunneling spectroscopy of three-dimensional 
normal-metal/Sr$_2$RuO$_4$ junctions enables to
determine the pairing symmetry of Sr$_2$RuO$_4$.
The differential conductances in the [001], [100], and [110]
direction 
have been obtained by the Blonder-Tinkham-Klapwijk theory. We have
considered three possible pairings, the spin-singlet $d_{zx}+id_{yz}$-wave, the spin-triplet helical
$p$-wave, and the spin-nematic $f_{xyz}$-wave pairings,
which are consistent with the NMR measurements.
Introducing the anisotropic effective-mass and the cutoff in the
momentum integration, the $\gamma$ band of Sr$_2$RuO$_4$ is modeled. 

Although the conductance spectra $\gns$ along the $c$-axis for the
chiral $d_{zx}+id_{yz}$-wave and $f_{xyz}$-wave are similar when there
is no spin-mixing (e.g., RSOI and the subdominant pair potential), 
the spectra is significantly modified by the spin-mixing depending on
the pairing symmetry:
$\gns$ for the $d_{zx}+id_{yz}$-wave is not qualitatively changed by
the spin-mixing, whereas the ZEP in the $\gns$ for the spin-nematic $f_{xyz}$-wave
is strongly suppressed. 
Comparing the calculated $\gns$ and the corresponding transport
experiments, we have concluded the spin-singlet $d_{zx}+id_{yz}$-wave
scenario does not explain the STM experiments, whereas the spin-nematic
$f_{xyz}$-wave and helical PW pairings do. 

We have also proposed that these two remaining candidates can be
distinguished without ambiguity by the in-plane Andreev spectroscopy.
The conductance spectra for the spin-nematic $f_{xyz}$-wave support a
ZEP and a V-shaped dip in the (100)- and (110)-interface junctions 
respectively, whereas those of the helical $p$-wave junction are
independent of the direction.


\begin{acknowledgements}
The authors would like to thank S.~Kashiwaya and S.~Kobayashi 
for useful discussions. 
S.-I.~S. is supported by Grant-in-Aid for JSPS Fellows (JSPS KAKENHI Grant Number JP19J02005). 
This work was supported by Grants-in-Aid from JSPS for Scientific
Research on Innovative Areas ``Topological Materials Science''
(KAKENHI Grant Numbers JP15H05851, JP15H05853, JP15H05855, and JP15K21717),
for Scientific Research (B) (KAKENHI Grant Number JP18H01176 and JP17H02922),
Japan-RFBR Bilateral Joint Research Projects/Seminars number 19-52-50026,  
and JSPS Core-to-Core Program (A. Advanced Research Networks). 
This work was supported by JST CREST (No: JPMJCR19T2), Japan.

\end{acknowledgements}

\bibliography{tsc07}

\begin{thebibliography}{76}%
\makeatletter
\providecommand \@ifxundefined [1]{%
 \@ifx{#1\undefined}
}%
\providecommand \@ifnum [1]{%
 \ifnum #1\expandafter \@firstoftwo
 \else \expandafter \@secondoftwo
 \fi
}%
\providecommand \@ifx [1]{%
 \ifx #1\expandafter \@firstoftwo
 \else \expandafter \@secondoftwo
 \fi
}%
\providecommand \natexlab [1]{#1}%
\providecommand \enquote  [1]{``#1''}%
\providecommand \bibnamefont  [1]{#1}%
\providecommand \bibfnamefont [1]{#1}%
\providecommand \citenamefont [1]{#1}%
\providecommand \href@noop [0]{\@secondoftwo}%
\providecommand \href [0]{\begingroup \@sanitize@url \@href}%
\providecommand \@href[1]{\@@startlink{#1}\@@href}%
\providecommand \@@href[1]{\endgroup#1\@@endlink}%
\providecommand \@sanitize@url [0]{\catcode `\\12\catcode `\$12\catcode
  `\&12\catcode `\#12\catcode `\^12\catcode `\_12\catcode `\%12\relax}%
\providecommand \@@startlink[1]{}%
\providecommand \@@endlink[0]{}%
\providecommand \url  [0]{\begingroup\@sanitize@url \@url }%
\providecommand \@url [1]{\endgroup\@href {#1}{\urlprefix }}%
\providecommand \urlprefix  [0]{URL }%
\providecommand \Eprint [0]{\href }%
\providecommand \doibase [0]{http://dx.doi.org/}%
\providecommand \selectlanguage [0]{\@gobble}%
\providecommand \bibinfo  [0]{\@secondoftwo}%
\providecommand \bibfield  [0]{\@secondoftwo}%
\providecommand \translation [1]{[#1]}%
\providecommand \BibitemOpen [0]{}%
\providecommand \bibitemStop [0]{}%
\providecommand \bibitemNoStop [0]{.\EOS\space}%
\providecommand \EOS [0]{\spacefactor3000\relax}%
\providecommand \BibitemShut  [1]{\csname bibitem#1\endcsname}%
\let\auto@bib@innerbib\@empty
\bibitem [{\citenamefont {Maeno}\ \emph {et~al.}(1994)\citenamefont {Maeno},
  \citenamefont {Hashimoto}, \citenamefont {Yoshida}, \citenamefont
  {Nishizaki}, \citenamefont {Fujita}, \citenamefont {Bednorz},\ and\
  \citenamefont {Lichtenberg}}]{Maeno_Nature_1994}%
  \BibitemOpen
  \bibfield  {author} {\bibinfo {author} {\bibfnamefont {Y.}~\bibnamefont
  {Maeno}}, \bibinfo {author} {\bibfnamefont {H.}~\bibnamefont {Hashimoto}},
  \bibinfo {author} {\bibfnamefont {K.}~\bibnamefont {Yoshida}}, \bibinfo
  {author} {\bibfnamefont {S.}~\bibnamefont {Nishizaki}}, \bibinfo {author}
  {\bibfnamefont {T.}~\bibnamefont {Fujita}}, \bibinfo {author} {\bibfnamefont
  {J.}~\bibnamefont {Bednorz}}, \ and\ \bibinfo {author} {\bibfnamefont
  {F.}~\bibnamefont {Lichtenberg}},\ }\href@noop {} {\bibfield  {journal}
  {\bibinfo  {journal} {Nature}\ }\textbf {\bibinfo {volume} {372}},\ \bibinfo
  {pages} {532} (\bibinfo {year} {1994})}\BibitemShut {NoStop}%
\bibitem [{\citenamefont {Mackenzie}\ and\ \citenamefont
  {Maeno}(2003)}]{Mackenzie_RMP_2003}%
  \BibitemOpen
  \bibfield  {author} {\bibinfo {author} {\bibfnamefont {A.~P.}\ \bibnamefont
  {Mackenzie}}\ and\ \bibinfo {author} {\bibfnamefont {Y.}~\bibnamefont
  {Maeno}},\ }\href {\doibase 10.1103/RevModPhys.75.657} {\bibfield  {journal}
  {\bibinfo  {journal} {Rev. Mod. Phys.}\ }\textbf {\bibinfo {volume} {75}},\
  \bibinfo {pages} {657} (\bibinfo {year} {2003})}\BibitemShut {NoStop}%
\bibitem [{\citenamefont {Maeno}\ \emph {et~al.}(2012)\citenamefont {Maeno},
  \citenamefont {Kittaka}, \citenamefont {Nomura}, \citenamefont {Yonezawa},\
  and\ \citenamefont {Ishida}}]{Maeno_JPSJ_2012}%
  \BibitemOpen
  \bibfield  {author} {\bibinfo {author} {\bibfnamefont {Y.}~\bibnamefont
  {Maeno}}, \bibinfo {author} {\bibfnamefont {S.}~\bibnamefont {Kittaka}},
  \bibinfo {author} {\bibfnamefont {T.}~\bibnamefont {Nomura}}, \bibinfo
  {author} {\bibfnamefont {S.}~\bibnamefont {Yonezawa}}, \ and\ \bibinfo
  {author} {\bibfnamefont {K.}~\bibnamefont {Ishida}},\ }\href@noop {}
  {\bibfield  {journal} {\bibinfo  {journal} {J. Phys. Soc. Jpn.}\ }\textbf
  {\bibinfo {volume} {81}},\ \bibinfo {pages} {011009} (\bibinfo {year}
  {2012})}\BibitemShut {NoStop}%
\bibitem [{\citenamefont {Rice}\ and\ \citenamefont
  {Sigrist}(1995)}]{Rice_1995}%
  \BibitemOpen
  \bibfield  {author} {\bibinfo {author} {\bibfnamefont {T.~M.}\ \bibnamefont
  {Rice}}\ and\ \bibinfo {author} {\bibfnamefont {M.}~\bibnamefont {Sigrist}},\
  }\href {http://stacks.iop.org/0953-8984/7/i=47/a=002} {\bibfield  {journal}
  {\bibinfo  {journal} {J. Phys.: Condens. Matter}\ }\textbf {\bibinfo {volume}
  {7}},\ \bibinfo {pages} {L643} (\bibinfo {year} {1995})}\BibitemShut
  {NoStop}%
\bibitem [{\citenamefont {Duffy}\ \emph {et~al.}(2000)\citenamefont {Duffy},
  \citenamefont {Hayden}, \citenamefont {Maeno}, \citenamefont {Mao},
  \citenamefont {Kulda},\ and\ \citenamefont {McIntyre}}]{Duffy_PRL_2000}%
  \BibitemOpen
  \bibfield  {author} {\bibinfo {author} {\bibfnamefont {J.~A.}\ \bibnamefont
  {Duffy}}, \bibinfo {author} {\bibfnamefont {S.~M.}\ \bibnamefont {Hayden}},
  \bibinfo {author} {\bibfnamefont {Y.}~\bibnamefont {Maeno}}, \bibinfo
  {author} {\bibfnamefont {Z.}~\bibnamefont {Mao}}, \bibinfo {author}
  {\bibfnamefont {J.}~\bibnamefont {Kulda}}, \ and\ \bibinfo {author}
  {\bibfnamefont {G.~J.}\ \bibnamefont {McIntyre}},\ }\href {\doibase
  10.1103/PhysRevLett.85.5412} {\bibfield  {journal} {\bibinfo  {journal}
  {Phys. Rev. Lett.}\ }\textbf {\bibinfo {volume} {85}},\ \bibinfo {pages}
  {5412} (\bibinfo {year} {2000})}\BibitemShut {NoStop}%
\bibitem [{\citenamefont {Jang}\ \emph {et~al.}(2011)\citenamefont {Jang},
  \citenamefont {Ferguson}, \citenamefont {Vakaryuk}, \citenamefont {Budakian},
  \citenamefont {Chung}, \citenamefont {Goldbart},\ and\ \citenamefont
  {Maeno}}]{Jang_Science_2011}%
  \BibitemOpen
  \bibfield  {author} {\bibinfo {author} {\bibfnamefont {J.}~\bibnamefont
  {Jang}}, \bibinfo {author} {\bibfnamefont {D.}~\bibnamefont {Ferguson}},
  \bibinfo {author} {\bibfnamefont {V.}~\bibnamefont {Vakaryuk}}, \bibinfo
  {author} {\bibfnamefont {R.}~\bibnamefont {Budakian}}, \bibinfo {author}
  {\bibfnamefont {S.}~\bibnamefont {Chung}}, \bibinfo {author} {\bibfnamefont
  {P.}~\bibnamefont {Goldbart}}, \ and\ \bibinfo {author} {\bibfnamefont
  {Y.}~\bibnamefont {Maeno}},\ }\href@noop {} {\bibfield  {journal} {\bibinfo
  {journal} {Science}\ }\textbf {\bibinfo {volume} {331}},\ \bibinfo {pages}
  {186} (\bibinfo {year} {2011})}\BibitemShut {NoStop}%
\bibitem [{\citenamefont {Yasui}\ \emph {et~al.}(2017)\citenamefont {Yasui},
  \citenamefont {Lahabi}, \citenamefont {Anwar}, \citenamefont {Nakamura},
  \citenamefont {Yonezawa}, \citenamefont {Terashima}, \citenamefont {Aarts},\
  and\ \citenamefont {Maeno}}]{Yasui_PRB_2017}%
  \BibitemOpen
  \bibfield  {author} {\bibinfo {author} {\bibfnamefont {Y.}~\bibnamefont
  {Yasui}}, \bibinfo {author} {\bibfnamefont {K.}~\bibnamefont {Lahabi}},
  \bibinfo {author} {\bibfnamefont {M.~S.}\ \bibnamefont {Anwar}}, \bibinfo
  {author} {\bibfnamefont {Y.}~\bibnamefont {Nakamura}}, \bibinfo {author}
  {\bibfnamefont {S.}~\bibnamefont {Yonezawa}}, \bibinfo {author}
  {\bibfnamefont {T.}~\bibnamefont {Terashima}}, \bibinfo {author}
  {\bibfnamefont {J.}~\bibnamefont {Aarts}}, \ and\ \bibinfo {author}
  {\bibfnamefont {Y.}~\bibnamefont {Maeno}},\ }\href {\doibase
  10.1103/PhysRevB.96.180507} {\bibfield  {journal} {\bibinfo  {journal} {Phys.
  Rev. B}\ }\textbf {\bibinfo {volume} {96}},\ \bibinfo {pages} {180507(R)}
  (\bibinfo {year} {2017})}\BibitemShut {NoStop}%
\bibitem [{\citenamefont {Laube}\ \emph {et~al.}(2000)\citenamefont {Laube},
  \citenamefont {Goll}, \citenamefont {L\"ohneysen}, \citenamefont
  {Fogelstr\"om},\ and\ \citenamefont {Lichtenberg}}]{Laube_PRL_2000}%
  \BibitemOpen
  \bibfield  {author} {\bibinfo {author} {\bibfnamefont {F.}~\bibnamefont
  {Laube}}, \bibinfo {author} {\bibfnamefont {G.}~\bibnamefont {Goll}},
  \bibinfo {author} {\bibfnamefont {H.~v.}\ \bibnamefont {L\"ohneysen}},
  \bibinfo {author} {\bibfnamefont {M.}~\bibnamefont {Fogelstr\"om}}, \ and\
  \bibinfo {author} {\bibfnamefont {F.}~\bibnamefont {Lichtenberg}},\ }\href
  {\doibase 10.1103/PhysRevLett.84.1595} {\bibfield  {journal} {\bibinfo
  {journal} {Phys. Rev. Lett.}\ }\textbf {\bibinfo {volume} {84}},\ \bibinfo
  {pages} {1595} (\bibinfo {year} {2000})}\BibitemShut {NoStop}%
\bibitem [{\citenamefont {Kashiwaya}\ \emph {et~al.}(2011)\citenamefont
  {Kashiwaya}, \citenamefont {Kashiwaya}, \citenamefont {Kambara},
  \citenamefont {Furuta}, \citenamefont {Yaguchi}, \citenamefont {Tanaka},\
  and\ \citenamefont {Maeno}}]{Kashiwaya_PRL_2011}%
  \BibitemOpen
  \bibfield  {author} {\bibinfo {author} {\bibfnamefont {S.}~\bibnamefont
  {Kashiwaya}}, \bibinfo {author} {\bibfnamefont {H.}~\bibnamefont
  {Kashiwaya}}, \bibinfo {author} {\bibfnamefont {H.}~\bibnamefont {Kambara}},
  \bibinfo {author} {\bibfnamefont {T.}~\bibnamefont {Furuta}}, \bibinfo
  {author} {\bibfnamefont {H.}~\bibnamefont {Yaguchi}}, \bibinfo {author}
  {\bibfnamefont {Y.}~\bibnamefont {Tanaka}}, \ and\ \bibinfo {author}
  {\bibfnamefont {Y.}~\bibnamefont {Maeno}},\ }\href {\doibase
  10.1103/PhysRevLett.107.077003} {\bibfield  {journal} {\bibinfo  {journal}
  {Phys. Rev. Lett.}\ }\textbf {\bibinfo {volume} {107}},\ \bibinfo {pages}
  {077003} (\bibinfo {year} {2011})}\BibitemShut {NoStop}%
\bibitem [{\citenamefont {Yamashiro}\ \emph
  {et~al.}(1998{\natexlab{a}})\citenamefont {Yamashiro}, \citenamefont
  {Tanaka},\ and\ \citenamefont {Kashiwaya}}]{Yamashiro_JPSJ_1998}%
  \BibitemOpen
  \bibfield  {author} {\bibinfo {author} {\bibfnamefont {M.}~\bibnamefont
  {Yamashiro}}, \bibinfo {author} {\bibfnamefont {Y.}~\bibnamefont {Tanaka}}, \
  and\ \bibinfo {author} {\bibfnamefont {S.}~\bibnamefont {Kashiwaya}},\ }\href
  {\doibase 10.1143/jpsj.67.3364} {\bibfield  {journal} {\bibinfo  {journal}
  {J. Phys. Soc. Jpn.}\ }\textbf {\bibinfo {volume} {67}},\ \bibinfo {pages}
  {3364} (\bibinfo {year} {1998}{\natexlab{a}})}\BibitemShut {NoStop}%
\bibitem [{\citenamefont {Jin}\ \emph {et~al.}(1999)\citenamefont {Jin},
  \citenamefont {Zadorozhny}, \citenamefont {Liu}, \citenamefont {Schlom},
  \citenamefont {Mori},\ and\ \citenamefont {Maeno}}]{Jin_PRB_1999}%
  \BibitemOpen
  \bibfield  {author} {\bibinfo {author} {\bibfnamefont {R.}~\bibnamefont
  {Jin}}, \bibinfo {author} {\bibfnamefont {Y.}~\bibnamefont {Zadorozhny}},
  \bibinfo {author} {\bibfnamefont {Y.}~\bibnamefont {Liu}}, \bibinfo {author}
  {\bibfnamefont {D.~G.}\ \bibnamefont {Schlom}}, \bibinfo {author}
  {\bibfnamefont {Y.}~\bibnamefont {Mori}}, \ and\ \bibinfo {author}
  {\bibfnamefont {Y.}~\bibnamefont {Maeno}},\ }\href {\doibase
  10.1103/PhysRevB.59.4433} {\bibfield  {journal} {\bibinfo  {journal} {Phys.
  Rev. B}\ }\textbf {\bibinfo {volume} {59}},\ \bibinfo {pages} {4433}
  (\bibinfo {year} {1999})}\BibitemShut {NoStop}%
\bibitem [{\citenamefont {Tanaka}\ \emph {et~al.}(2009)\citenamefont {Tanaka},
  \citenamefont {Yokoyama}, \citenamefont {Balatsky},\ and\ \citenamefont
  {Nagaosa}}]{TYBN09}%
  \BibitemOpen
  \bibfield  {author} {\bibinfo {author} {\bibfnamefont {Y.}~\bibnamefont
  {Tanaka}}, \bibinfo {author} {\bibfnamefont {T.}~\bibnamefont {Yokoyama}},
  \bibinfo {author} {\bibfnamefont {A.~V.}\ \bibnamefont {Balatsky}}, \ and\
  \bibinfo {author} {\bibfnamefont {N.}~\bibnamefont {Nagaosa}},\ }\href@noop
  {} {\bibfield  {journal} {\bibinfo  {journal} {Phys. Rev. B}\ }\textbf
  {\bibinfo {volume} {79}},\ \bibinfo {pages} {060505(R)} (\bibinfo {year}
  {2009})}\BibitemShut {NoStop}%
\bibitem [{\citenamefont {Wu}\ and\ \citenamefont {Samokhin}(2010)}]{Samokhin}%
  \BibitemOpen
  \bibfield  {author} {\bibinfo {author} {\bibfnamefont {S.}~\bibnamefont
  {Wu}}\ and\ \bibinfo {author} {\bibfnamefont {K.~V.}\ \bibnamefont
  {Samokhin}},\ }\href@noop {} {\bibfield  {journal} {\bibinfo  {journal}
  {Phys. Rev. B}\ }\textbf {\bibinfo {volume} {81}},\ \bibinfo {pages} {214506}
  (\bibinfo {year} {2010})}\BibitemShut {NoStop}%
\bibitem [{\citenamefont {Anwar}\ \emph {et~al.}(2016)\citenamefont {Anwar},
  \citenamefont {Lee}, \citenamefont {Ishiguro}, \citenamefont {Sugimoto},
  \citenamefont {Tano}, \citenamefont {Kang}, \citenamefont {Shin},
  \citenamefont {Yonezawa}, \citenamefont {Manske}, \citenamefont {Takayanagi}
  \emph {et~al.}}]{Anwar_NatComn_2016}%
  \BibitemOpen
  \bibfield  {author} {\bibinfo {author} {\bibfnamefont {M.}~\bibnamefont
  {Anwar}}, \bibinfo {author} {\bibfnamefont {S.}~\bibnamefont {Lee}}, \bibinfo
  {author} {\bibfnamefont {R.}~\bibnamefont {Ishiguro}}, \bibinfo {author}
  {\bibfnamefont {Y.}~\bibnamefont {Sugimoto}}, \bibinfo {author}
  {\bibfnamefont {Y.}~\bibnamefont {Tano}}, \bibinfo {author} {\bibfnamefont
  {S.}~\bibnamefont {Kang}}, \bibinfo {author} {\bibfnamefont {Y.}~\bibnamefont
  {Shin}}, \bibinfo {author} {\bibfnamefont {S.}~\bibnamefont {Yonezawa}},
  \bibinfo {author} {\bibfnamefont {D.}~\bibnamefont {Manske}}, \bibinfo
  {author} {\bibfnamefont {H.}~\bibnamefont {Takayanagi}},  \emph {et~al.},\
  }\href@noop {} {\bibfield  {journal} {\bibinfo  {journal} {Nat. Commun.}\
  }\textbf {\bibinfo {volume} {7}},\ \bibinfo {pages} {13220} (\bibinfo {year}
  {2016})}\BibitemShut {NoStop}%
\bibitem [{\citenamefont {Olde~Olthof}\ \emph {et~al.}(2018)\citenamefont
  {Olde~Olthof}, \citenamefont {Suzuki}, \citenamefont {Golubov}, \citenamefont
  {Kunieda}, \citenamefont {Yonezawa}, \citenamefont {Maeno},\ and\
  \citenamefont {Tanaka}}]{OldeOlthof_PRB_2018}%
  \BibitemOpen
  \bibfield  {author} {\bibinfo {author} {\bibfnamefont {L.~A.~B.}\
  \bibnamefont {Olde~Olthof}}, \bibinfo {author} {\bibfnamefont {S.-I.}\
  \bibnamefont {Suzuki}}, \bibinfo {author} {\bibfnamefont {A.~A.}\
  \bibnamefont {Golubov}}, \bibinfo {author} {\bibfnamefont {M.}~\bibnamefont
  {Kunieda}}, \bibinfo {author} {\bibfnamefont {S.}~\bibnamefont {Yonezawa}},
  \bibinfo {author} {\bibfnamefont {Y.}~\bibnamefont {Maeno}}, \ and\ \bibinfo
  {author} {\bibfnamefont {Y.}~\bibnamefont {Tanaka}},\ }\href {\doibase
  10.1103/PhysRevB.98.014508} {\bibfield  {journal} {\bibinfo  {journal} {Phys.
  Rev. B}\ }\textbf {\bibinfo {volume} {98}},\ \bibinfo {pages} {014508}
  (\bibinfo {year} {2018})}\BibitemShut {NoStop}%
\bibitem [{\citenamefont {Ishida}\ \emph {et~al.}(1998)\citenamefont {Ishida},
  \citenamefont {Mukuda}, \citenamefont {Kitaoka}, \citenamefont {Asayama},
  \citenamefont {Mao}, \citenamefont {Mori},\ and\ \citenamefont
  {Maeno}}]{Ishida_Nature_1998}%
  \BibitemOpen
  \bibfield  {author} {\bibinfo {author} {\bibfnamefont {K.}~\bibnamefont
  {Ishida}}, \bibinfo {author} {\bibfnamefont {H.}~\bibnamefont {Mukuda}},
  \bibinfo {author} {\bibfnamefont {Y.}~\bibnamefont {Kitaoka}}, \bibinfo
  {author} {\bibfnamefont {K.}~\bibnamefont {Asayama}}, \bibinfo {author}
  {\bibfnamefont {Z.}~\bibnamefont {Mao}}, \bibinfo {author} {\bibfnamefont
  {Y.}~\bibnamefont {Mori}}, \ and\ \bibinfo {author} {\bibfnamefont
  {Y.}~\bibnamefont {Maeno}},\ }\href@noop {} {\bibfield  {journal} {\bibinfo
  {journal} {Nature}\ }\textbf {\bibinfo {volume} {396}},\ \bibinfo {pages}
  {658} (\bibinfo {year} {1998})}\BibitemShut {NoStop}%
\bibitem [{Nom()}]{Nomura}%
  \BibitemOpen
  \href@noop {} {\ }\bibinfo {note} {T.~Nomura and K.~Yamada, J. Phys. Soc.
  Jpn. \textbf{69}, 3678 (2000). \textit{Ibid.} \textbf{71}, 404 (2002).
  \textit{Ibid.} \textbf{71}, 1993 (2002). \textit{Ibid.} \textbf{74}, 1818
  (2004).}\BibitemShut {Stop}%
\bibitem [{\citenamefont {Nomura}\ \emph {et~al.}(2008)\citenamefont {Nomura},
  \citenamefont {Hirashima},\ and\ \citenamefont {Yamada}}]{Nomura08}%
  \BibitemOpen
  \bibfield  {author} {\bibinfo {author} {\bibfnamefont {T.}~\bibnamefont
  {Nomura}}, \bibinfo {author} {\bibfnamefont {D.~S.}\ \bibnamefont
  {Hirashima}}, \ and\ \bibinfo {author} {\bibfnamefont {K.}~\bibnamefont
  {Yamada}},\ }\href@noop {} {\bibfield  {journal} {\bibinfo  {journal} {J.
  Phys. Soc. Jpn.}\ }\textbf {\bibinfo {volume} {77}},\ \bibinfo {pages}
  {024701} (\bibinfo {year} {2008})}\BibitemShut {NoStop}%
\bibitem [{\citenamefont {Yanase}\ and\ \citenamefont {Ogata}(2003)}]{Yanase}%
  \BibitemOpen
  \bibfield  {author} {\bibinfo {author} {\bibfnamefont {Y.}~\bibnamefont
  {Yanase}}\ and\ \bibinfo {author} {\bibfnamefont {M.}~\bibnamefont {Ogata}},\
  }\href@noop {} {\bibfield  {journal} {\bibinfo  {journal} {J. Phys. Soc.
  Jpn.}\ }\textbf {\bibinfo {volume} {72}},\ \bibinfo {pages} {673} (\bibinfo
  {year} {2003})}\BibitemShut {NoStop}%
\bibitem [{\citenamefont {Raghu}\ \emph {et~al.}(2010)\citenamefont {Raghu},
  \citenamefont {Kapitulnik},\ and\ \citenamefont {Kivelson}}]{Raghu}%
  \BibitemOpen
  \bibfield  {author} {\bibinfo {author} {\bibfnamefont {S.}~\bibnamefont
  {Raghu}}, \bibinfo {author} {\bibfnamefont {A.}~\bibnamefont {Kapitulnik}}, \
  and\ \bibinfo {author} {\bibfnamefont {S.~A.}\ \bibnamefont {Kivelson}},\
  }\href@noop {} {\bibfield  {journal} {\bibinfo  {journal} {Phys. Rev. Lett.}\
  }\textbf {\bibinfo {volume} {105}},\ \bibinfo {pages} {136401} (\bibinfo
  {year} {2010})}\BibitemShut {NoStop}%
\bibitem [{\citenamefont {Sato}\ and\ \citenamefont {Kohmoto}()}]{Kohmoto}%
  \BibitemOpen
  \bibfield  {author} {\bibinfo {author} {\bibfnamefont {M.}~\bibnamefont
  {Sato}}\ and\ \bibinfo {author} {\bibfnamefont {M.}~\bibnamefont {Kohmoto}},\
  }\href@noop {} {\bibfield  {journal} {\bibinfo  {journal} {J. Phys. Soc.
  Jpn.}\ }\textbf {\bibinfo {volume} {69}},\ \bibinfo {pages}
  {3505}}\BibitemShut {NoStop}%
\bibitem [{\citenamefont {Kuroki}\ \emph {et~al.}(2001)\citenamefont {Kuroki},
  \citenamefont {Ogata}, \citenamefont {Arita},\ and\ \citenamefont
  {Aoki}}]{Kuroki}%
  \BibitemOpen
  \bibfield  {author} {\bibinfo {author} {\bibfnamefont {K.}~\bibnamefont
  {Kuroki}}, \bibinfo {author} {\bibfnamefont {M.}~\bibnamefont {Ogata}},
  \bibinfo {author} {\bibfnamefont {R.}~\bibnamefont {Arita}}, \ and\ \bibinfo
  {author} {\bibfnamefont {H.}~\bibnamefont {Aoki}},\ }\href@noop {} {\bibfield
   {journal} {\bibinfo  {journal} {Phys. Rev. B}\ }\textbf {\bibinfo {volume}
  {63}},\ \bibinfo {pages} {060506} (\bibinfo {year} {2001})}\BibitemShut
  {NoStop}%
\bibitem [{\citenamefont {Takimoto}(2000)}]{Takimoto}%
  \BibitemOpen
  \bibfield  {author} {\bibinfo {author} {\bibfnamefont {T.}~\bibnamefont
  {Takimoto}},\ }\href@noop {} {\bibfield  {journal} {\bibinfo  {journal}
  {Phys. Rev. B}\ }\textbf {\bibinfo {volume} {62}},\ \bibinfo {pages} {R14641}
  (\bibinfo {year} {2000})}\BibitemShut {NoStop}%
\bibitem [{\citenamefont {Tsuchiizu}\ \emph {et~al.}(2015)\citenamefont
  {Tsuchiizu}, \citenamefont {Yamakawa}, \citenamefont {Onari}, \citenamefont
  {Ohno},\ and\ \citenamefont {Kontani}}]{Tsuchiizu}%
  \BibitemOpen
  \bibfield  {author} {\bibinfo {author} {\bibfnamefont {M.}~\bibnamefont
  {Tsuchiizu}}, \bibinfo {author} {\bibfnamefont {Y.}~\bibnamefont {Yamakawa}},
  \bibinfo {author} {\bibfnamefont {S.}~\bibnamefont {Onari}}, \bibinfo
  {author} {\bibfnamefont {Y.}~\bibnamefont {Ohno}}, \ and\ \bibinfo {author}
  {\bibfnamefont {H.}~\bibnamefont {Kontani}},\ }\href {\doibase
  10.1103/PhysRevB.91.155103} {\bibfield  {journal} {\bibinfo  {journal} {Phys.
  Rev. B}\ }\textbf {\bibinfo {volume} {91}},\ \bibinfo {pages} {155103}
  (\bibinfo {year} {2015})}\BibitemShut {NoStop}%
\bibitem [{\citenamefont {Pustogow}\ \emph {et~al.}(2019)\citenamefont
  {Pustogow}, \citenamefont {Luo}, \citenamefont {Chronister}, \citenamefont
  {Su}, \citenamefont {Sokolov}, \citenamefont {Jerzembeck}, \citenamefont
  {Mackenzie}, \citenamefont {Hicks}, \citenamefont {Kikugawa}, \citenamefont
  {Raghu} \emph {et~al.}}]{Pustogow_arXiv_2019}%
  \BibitemOpen
  \bibfield  {author} {\bibinfo {author} {\bibfnamefont {A.}~\bibnamefont
  {Pustogow}}, \bibinfo {author} {\bibfnamefont {Y.}~\bibnamefont {Luo}},
  \bibinfo {author} {\bibfnamefont {A.}~\bibnamefont {Chronister}}, \bibinfo
  {author} {\bibfnamefont {Y.-S.}\ \bibnamefont {Su}}, \bibinfo {author}
  {\bibfnamefont {D.}~\bibnamefont {Sokolov}}, \bibinfo {author} {\bibfnamefont
  {F.}~\bibnamefont {Jerzembeck}}, \bibinfo {author} {\bibfnamefont
  {A.}~\bibnamefont {Mackenzie}}, \bibinfo {author} {\bibfnamefont
  {C.}~\bibnamefont {Hicks}}, \bibinfo {author} {\bibfnamefont
  {N.}~\bibnamefont {Kikugawa}}, \bibinfo {author} {\bibfnamefont
  {S.}~\bibnamefont {Raghu}},  \emph {et~al.},\ }\href
  {https://doi.org/10.1038/s41586-019-1596-2} {\bibfield  {journal} {\bibinfo
  {journal} {Nature}\ }\textbf {\bibinfo {volume} {574}},\ \bibinfo {pages}
  {72} (\bibinfo {year} {2019})}\BibitemShut {NoStop}%
\bibitem [{\citenamefont {Ishida}\ \emph {et~al.}(2019)\citenamefont {Ishida},
  \citenamefont {Manago},\ and\ \citenamefont {Maeno}}]{Ishida_arXiv_2019}%
  \BibitemOpen
  \bibfield  {author} {\bibinfo {author} {\bibfnamefont {K.}~\bibnamefont
  {Ishida}}, \bibinfo {author} {\bibfnamefont {M.}~\bibnamefont {Manago}}, \
  and\ \bibinfo {author} {\bibfnamefont {Y.}~\bibnamefont {Maeno}},\
  }\href@noop {} {\bibfield  {journal} {\bibinfo  {journal} {arXiv:1907.12236}\
  } (\bibinfo {year} {2019})}\BibitemShut {NoStop}%
\bibitem [{\citenamefont {Yonezawa}\ \emph {et~al.}(2013)\citenamefont
  {Yonezawa}, \citenamefont {Kajikawa},\ and\ \citenamefont
  {Maeno}}]{Yonezawa_PRL_2013}%
  \BibitemOpen
  \bibfield  {author} {\bibinfo {author} {\bibfnamefont {S.}~\bibnamefont
  {Yonezawa}}, \bibinfo {author} {\bibfnamefont {T.}~\bibnamefont {Kajikawa}},
  \ and\ \bibinfo {author} {\bibfnamefont {Y.}~\bibnamefont {Maeno}},\ }\href
  {\doibase 10.1103/PhysRevLett.110.077003} {\bibfield  {journal} {\bibinfo
  {journal} {Phys. Rev. Lett.}\ }\textbf {\bibinfo {volume} {110}},\ \bibinfo
  {pages} {077003} (\bibinfo {year} {2013})}\BibitemShut {NoStop}%
\bibitem [{\citenamefont {Yonezawa}\ \emph {et~al.}(2014)\citenamefont
  {Yonezawa}, \citenamefont {Kajikawa},\ and\ \citenamefont
  {Maeno}}]{yonezawa_JPSJ_2014}%
  \BibitemOpen
  \bibfield  {author} {\bibinfo {author} {\bibfnamefont {S.}~\bibnamefont
  {Yonezawa}}, \bibinfo {author} {\bibfnamefont {T.}~\bibnamefont {Kajikawa}},
  \ and\ \bibinfo {author} {\bibfnamefont {Y.}~\bibnamefont {Maeno}},\ }\href
  {https://doi.org/10.7566/JPSJ.83.083706} {\bibfield  {journal} {\bibinfo
  {journal} {J. Phys. Soc. Jpn.}\ }\textbf {\bibinfo {volume} {83}},\ \bibinfo
  {pages} {083706} (\bibinfo {year} {2014})}\BibitemShut {NoStop}%
\bibitem [{\citenamefont {Kittaka}\ \emph {et~al.}(2014)\citenamefont
  {Kittaka}, \citenamefont {Kasahara}, \citenamefont {Sakakibara},
  \citenamefont {Shibata}, \citenamefont {Yonezawa}, \citenamefont {Maeno},
  \citenamefont {Tenya},\ and\ \citenamefont {Machida}}]{Kittaka_PRB_2014}%
  \BibitemOpen
  \bibfield  {author} {\bibinfo {author} {\bibfnamefont {S.}~\bibnamefont
  {Kittaka}}, \bibinfo {author} {\bibfnamefont {A.}~\bibnamefont {Kasahara}},
  \bibinfo {author} {\bibfnamefont {T.}~\bibnamefont {Sakakibara}}, \bibinfo
  {author} {\bibfnamefont {D.}~\bibnamefont {Shibata}}, \bibinfo {author}
  {\bibfnamefont {S.}~\bibnamefont {Yonezawa}}, \bibinfo {author}
  {\bibfnamefont {Y.}~\bibnamefont {Maeno}}, \bibinfo {author} {\bibfnamefont
  {K.}~\bibnamefont {Tenya}}, \ and\ \bibinfo {author} {\bibfnamefont
  {K.}~\bibnamefont {Machida}},\ }\href {\doibase 10.1103/PhysRevB.90.220502}
  {\bibfield  {journal} {\bibinfo  {journal} {Phys. Rev. B}\ }\textbf {\bibinfo
  {volume} {90}},\ \bibinfo {pages} {220502} (\bibinfo {year}
  {2014})}\BibitemShut {NoStop}%
\bibitem [{\citenamefont {Luke}\ \emph {et~al.}(1998)\citenamefont {Luke},
  \citenamefont {Fudamoto}, \citenamefont {Kojima}, \citenamefont {Larkin},
  \citenamefont {Merrin}, \citenamefont {Nachumi}, \citenamefont {Uemura},
  \citenamefont {Maeno}, \citenamefont {Mao}, \citenamefont {Mori} \emph
  {et~al.}}]{Luke_Nature_1998}%
  \BibitemOpen
  \bibfield  {author} {\bibinfo {author} {\bibfnamefont {G.~M.}\ \bibnamefont
  {Luke}}, \bibinfo {author} {\bibfnamefont {Y.}~\bibnamefont {Fudamoto}},
  \bibinfo {author} {\bibfnamefont {K.}~\bibnamefont {Kojima}}, \bibinfo
  {author} {\bibfnamefont {M.}~\bibnamefont {Larkin}}, \bibinfo {author}
  {\bibfnamefont {J.}~\bibnamefont {Merrin}}, \bibinfo {author} {\bibfnamefont
  {B.}~\bibnamefont {Nachumi}}, \bibinfo {author} {\bibfnamefont
  {Y.}~\bibnamefont {Uemura}}, \bibinfo {author} {\bibfnamefont
  {Y.}~\bibnamefont {Maeno}}, \bibinfo {author} {\bibfnamefont
  {Z.}~\bibnamefont {Mao}}, \bibinfo {author} {\bibfnamefont {Y.}~\bibnamefont
  {Mori}},  \emph {et~al.},\ }\href@noop {} {\bibfield  {journal} {\bibinfo
  {journal} {Nature}\ }\textbf {\bibinfo {volume} {394}},\ \bibinfo {pages}
  {558} (\bibinfo {year} {1998})}\BibitemShut {NoStop}%
\bibitem [{\citenamefont {Xia}\ \emph {et~al.}(2006)\citenamefont {Xia},
  \citenamefont {Maeno}, \citenamefont {Beyersdorf}, \citenamefont {Fejer},\
  and\ \citenamefont {Kapitulnik}}]{Xia_PRL_2006}%
  \BibitemOpen
  \bibfield  {author} {\bibinfo {author} {\bibfnamefont {J.}~\bibnamefont
  {Xia}}, \bibinfo {author} {\bibfnamefont {Y.}~\bibnamefont {Maeno}}, \bibinfo
  {author} {\bibfnamefont {P.~T.}\ \bibnamefont {Beyersdorf}}, \bibinfo
  {author} {\bibfnamefont {M.~M.}\ \bibnamefont {Fejer}}, \ and\ \bibinfo
  {author} {\bibfnamefont {A.}~\bibnamefont {Kapitulnik}},\ }\href {\doibase
  10.1103/PhysRevLett.97.167002} {\bibfield  {journal} {\bibinfo  {journal}
  {Phys. Rev. Lett.}\ }\textbf {\bibinfo {volume} {97}},\ \bibinfo {pages}
  {167002} (\bibinfo {year} {2006})}\BibitemShut {NoStop}%
\bibitem [{\citenamefont {Lupien}\ \emph {et~al.}(2001)\citenamefont {Lupien},
  \citenamefont {MacFarlane}, \citenamefont {Proust}, \citenamefont
  {Taillefer}, \citenamefont {Mao},\ and\ \citenamefont
  {Maeno}}]{Lupien_PRL_2001}%
  \BibitemOpen
  \bibfield  {author} {\bibinfo {author} {\bibfnamefont {C.}~\bibnamefont
  {Lupien}}, \bibinfo {author} {\bibfnamefont {W.~A.}\ \bibnamefont
  {MacFarlane}}, \bibinfo {author} {\bibfnamefont {C.}~\bibnamefont {Proust}},
  \bibinfo {author} {\bibfnamefont {L.}~\bibnamefont {Taillefer}}, \bibinfo
  {author} {\bibfnamefont {Z.~Q.}\ \bibnamefont {Mao}}, \ and\ \bibinfo
  {author} {\bibfnamefont {Y.}~\bibnamefont {Maeno}},\ }\href {\doibase
  10.1103/PhysRevLett.86.5986} {\bibfield  {journal} {\bibinfo  {journal}
  {Phys. Rev. Lett.}\ }\textbf {\bibinfo {volume} {86}},\ \bibinfo {pages}
  {5986} (\bibinfo {year} {2001})}\BibitemShut {NoStop}%
\bibitem [{\citenamefont {Okuda}\ \emph {et~al.}(2002)\citenamefont {Okuda},
  \citenamefont {Suzuki}, \citenamefont {Mao}, \citenamefont {Maeno},\ and\
  \citenamefont {Fujita}}]{Okuda_JPSJ_2002}%
  \BibitemOpen
  \bibfield  {author} {\bibinfo {author} {\bibfnamefont {N.}~\bibnamefont
  {Okuda}}, \bibinfo {author} {\bibfnamefont {T.}~\bibnamefont {Suzuki}},
  \bibinfo {author} {\bibfnamefont {Z.}~\bibnamefont {Mao}}, \bibinfo {author}
  {\bibfnamefont {Y.}~\bibnamefont {Maeno}}, \ and\ \bibinfo {author}
  {\bibfnamefont {T.}~\bibnamefont {Fujita}},\ }\href
  {https://doi.org/10.1143/JPSJ.71.1134} {\bibfield  {journal} {\bibinfo
  {journal} {J. Phys. Soc. Jpn.}\ }\textbf {\bibinfo {volume} {71}},\ \bibinfo
  {pages} {1134} (\bibinfo {year} {2002})}\BibitemShut {NoStop}%
\bibitem [{\citenamefont {Matsumoto}\ and\ \citenamefont
  {Sigrist}(1999)}]{Matsumoto_JPSJ_1999}%
  \BibitemOpen
  \bibfield  {author} {\bibinfo {author} {\bibfnamefont {M.}~\bibnamefont
  {Matsumoto}}\ and\ \bibinfo {author} {\bibfnamefont {M.}~\bibnamefont
  {Sigrist}},\ }\href@noop {} {\bibfield  {journal} {\bibinfo  {journal} {J.
  Phys. Soc. Jpn.}\ }\textbf {\bibinfo {volume} {68}},\ \bibinfo {pages} {994}
  (\bibinfo {year} {1999})}\BibitemShut {NoStop}%
\bibitem [{\citenamefont {Furusaki}\ \emph {et~al.}(2001)\citenamefont
  {Furusaki}, \citenamefont {Matsumoto},\ and\ \citenamefont
  {Sigrist}}]{Furusaki_PRB_2001}%
  \BibitemOpen
  \bibfield  {author} {\bibinfo {author} {\bibfnamefont {A.}~\bibnamefont
  {Furusaki}}, \bibinfo {author} {\bibfnamefont {M.}~\bibnamefont {Matsumoto}},
  \ and\ \bibinfo {author} {\bibfnamefont {M.}~\bibnamefont {Sigrist}},\ }\href
  {\doibase 10.1103/PhysRevB.64.054514} {\bibfield  {journal} {\bibinfo
  {journal} {Phys. Rev. B}\ }\textbf {\bibinfo {volume} {64}},\ \bibinfo
  {pages} {054514} (\bibinfo {year} {2001})}\BibitemShut {NoStop}%
\bibitem [{\citenamefont {Bj\"ornsson}\ \emph {et~al.}(2005)\citenamefont
  {Bj\"ornsson}, \citenamefont {Maeno}, \citenamefont {Huber},\ and\
  \citenamefont {Moler}}]{Bjornsson_PRB_2005}%
  \BibitemOpen
  \bibfield  {author} {\bibinfo {author} {\bibfnamefont {P.~G.}\ \bibnamefont
  {Bj\"ornsson}}, \bibinfo {author} {\bibfnamefont {Y.}~\bibnamefont {Maeno}},
  \bibinfo {author} {\bibfnamefont {M.~E.}\ \bibnamefont {Huber}}, \ and\
  \bibinfo {author} {\bibfnamefont {K.~A.}\ \bibnamefont {Moler}},\ }\href
  {\doibase 10.1103/PhysRevB.72.012504} {\bibfield  {journal} {\bibinfo
  {journal} {Phys. Rev. B}\ }\textbf {\bibinfo {volume} {72}},\ \bibinfo
  {pages} {012504} (\bibinfo {year} {2005})}\BibitemShut {NoStop}%
\bibitem [{\citenamefont {Kirtley}\ \emph {et~al.}(2007)\citenamefont
  {Kirtley}, \citenamefont {Kallin}, \citenamefont {Hicks}, \citenamefont
  {Kim}, \citenamefont {Liu}, \citenamefont {Moler}, \citenamefont {Maeno},\
  and\ \citenamefont {Nelson}}]{Kirtley_PRB_2007}%
  \BibitemOpen
  \bibfield  {author} {\bibinfo {author} {\bibfnamefont {J.~R.}\ \bibnamefont
  {Kirtley}}, \bibinfo {author} {\bibfnamefont {C.}~\bibnamefont {Kallin}},
  \bibinfo {author} {\bibfnamefont {C.~W.}\ \bibnamefont {Hicks}}, \bibinfo
  {author} {\bibfnamefont {E.-A.}\ \bibnamefont {Kim}}, \bibinfo {author}
  {\bibfnamefont {Y.}~\bibnamefont {Liu}}, \bibinfo {author} {\bibfnamefont
  {K.~A.}\ \bibnamefont {Moler}}, \bibinfo {author} {\bibfnamefont
  {Y.}~\bibnamefont {Maeno}}, \ and\ \bibinfo {author} {\bibfnamefont {K.~D.}\
  \bibnamefont {Nelson}},\ }\href {\doibase 10.1103/PhysRevB.76.014526}
  {\bibfield  {journal} {\bibinfo  {journal} {Phys. Rev. B}\ }\textbf {\bibinfo
  {volume} {76}},\ \bibinfo {pages} {014526} (\bibinfo {year}
  {2007})}\BibitemShut {NoStop}%
\bibitem [{\citenamefont {Hicks}\ \emph {et~al.}(2010)\citenamefont {Hicks},
  \citenamefont {Kirtley}, \citenamefont {Lippman}, \citenamefont {Koshnick},
  \citenamefont {Huber}, \citenamefont {Maeno}, \citenamefont {Yuhasz},
  \citenamefont {Maple},\ and\ \citenamefont {Moler}}]{Hicks_PRB_2010}%
  \BibitemOpen
  \bibfield  {author} {\bibinfo {author} {\bibfnamefont {C.~W.}\ \bibnamefont
  {Hicks}}, \bibinfo {author} {\bibfnamefont {J.~R.}\ \bibnamefont {Kirtley}},
  \bibinfo {author} {\bibfnamefont {T.~M.}\ \bibnamefont {Lippman}}, \bibinfo
  {author} {\bibfnamefont {N.~C.}\ \bibnamefont {Koshnick}}, \bibinfo {author}
  {\bibfnamefont {M.~E.}\ \bibnamefont {Huber}}, \bibinfo {author}
  {\bibfnamefont {Y.}~\bibnamefont {Maeno}}, \bibinfo {author} {\bibfnamefont
  {W.~M.}\ \bibnamefont {Yuhasz}}, \bibinfo {author} {\bibfnamefont {M.~B.}\
  \bibnamefont {Maple}}, \ and\ \bibinfo {author} {\bibfnamefont {K.~A.}\
  \bibnamefont {Moler}},\ }\href {\doibase 10.1103/PhysRevB.81.214501}
  {\bibfield  {journal} {\bibinfo  {journal} {Phys. Rev. B}\ }\textbf {\bibinfo
  {volume} {81}},\ \bibinfo {pages} {214501} (\bibinfo {year}
  {2010})}\BibitemShut {NoStop}%
\bibitem [{\citenamefont {Kallin}(2012)}]{Kallin_2012}%
  \BibitemOpen
  \bibfield  {author} {\bibinfo {author} {\bibfnamefont {C.}~\bibnamefont
  {Kallin}},\ }\href@noop {} {\bibfield  {journal} {\bibinfo  {journal}
  {Reports on Progress in Physics}\ }\textbf {\bibinfo {volume} {75}},\
  \bibinfo {pages} {042501} (\bibinfo {year} {2012})}\BibitemShut {NoStop}%
\bibitem [{\citenamefont {Curran}\ \emph {et~al.}(2014)\citenamefont {Curran},
  \citenamefont {Bending}, \citenamefont {Desoky}, \citenamefont {Gibbs},
  \citenamefont {Lee},\ and\ \citenamefont {Mackenzie}}]{Curran_PRB_2014}%
  \BibitemOpen
  \bibfield  {author} {\bibinfo {author} {\bibfnamefont {P.~J.}\ \bibnamefont
  {Curran}}, \bibinfo {author} {\bibfnamefont {S.~J.}\ \bibnamefont {Bending}},
  \bibinfo {author} {\bibfnamefont {W.~M.}\ \bibnamefont {Desoky}}, \bibinfo
  {author} {\bibfnamefont {A.~S.}\ \bibnamefont {Gibbs}}, \bibinfo {author}
  {\bibfnamefont {S.~L.}\ \bibnamefont {Lee}}, \ and\ \bibinfo {author}
  {\bibfnamefont {A.~P.}\ \bibnamefont {Mackenzie}},\ }\href {\doibase
  10.1103/PhysRevB.89.144504} {\bibfield  {journal} {\bibinfo  {journal} {Phys.
  Rev. B}\ }\textbf {\bibinfo {volume} {89}},\ \bibinfo {pages} {144504}
  (\bibinfo {year} {2014})}\BibitemShut {NoStop}%
\bibitem [{\citenamefont {Suzuki}\ and\ \citenamefont {Asano}(2016)}]{SIS3}%
  \BibitemOpen
  \bibfield  {author} {\bibinfo {author} {\bibfnamefont {S.-I.}\ \bibnamefont
  {Suzuki}}\ and\ \bibinfo {author} {\bibfnamefont {Y.}~\bibnamefont {Asano}},\
  }\href {\doibase 10.1103/PhysRevB.94.155302} {\bibfield  {journal} {\bibinfo
  {journal} {Phys. Rev. B}\ }\textbf {\bibinfo {volume} {94}},\ \bibinfo
  {pages} {155302} (\bibinfo {year} {2016})}\BibitemShut {NoStop}%
\bibitem [{\citenamefont {Bakurskiy}\ \emph {et~al.}(2017)\citenamefont
  {Bakurskiy}, \citenamefont {Klenov}, \citenamefont {Soloviev}, \citenamefont
  {Kupriyanov},\ and\ \citenamefont {Golubov}}]{Bakurskiy_2017}%
  \BibitemOpen
  \bibfield  {author} {\bibinfo {author} {\bibfnamefont {S.~V.}\ \bibnamefont
  {Bakurskiy}}, \bibinfo {author} {\bibfnamefont {N.~V.}\ \bibnamefont
  {Klenov}}, \bibinfo {author} {\bibfnamefont {I.~I.}\ \bibnamefont
  {Soloviev}}, \bibinfo {author} {\bibfnamefont {M.~Y.}\ \bibnamefont
  {Kupriyanov}}, \ and\ \bibinfo {author} {\bibfnamefont {A.~A.}\ \bibnamefont
  {Golubov}},\ }\href {\doibase 10.1088/1361-6668/aa5f3d} {\bibfield  {journal}
  {\bibinfo  {journal} {Supercond. Sci. Technol.}\ }\textbf {\bibinfo {volume}
  {30}},\ \bibinfo {pages} {044005} (\bibinfo {year} {2017})}\BibitemShut
  {NoStop}%
\bibitem [{\citenamefont {Kawai}\ \emph {et~al.}(2017)\citenamefont {Kawai},
  \citenamefont {Yada}, \citenamefont {Tanaka}, \citenamefont {Asano},
  \citenamefont {Golubov},\ and\ \citenamefont {Kashiwaya}}]{Kawai_PRB_2017}%
  \BibitemOpen
  \bibfield  {author} {\bibinfo {author} {\bibfnamefont {K.}~\bibnamefont
  {Kawai}}, \bibinfo {author} {\bibfnamefont {K.}~\bibnamefont {Yada}},
  \bibinfo {author} {\bibfnamefont {Y.}~\bibnamefont {Tanaka}}, \bibinfo
  {author} {\bibfnamefont {Y.}~\bibnamefont {Asano}}, \bibinfo {author}
  {\bibfnamefont {A.~A.}\ \bibnamefont {Golubov}}, \ and\ \bibinfo {author}
  {\bibfnamefont {S.}~\bibnamefont {Kashiwaya}},\ }\href {\doibase
  10.1103/PhysRevB.95.174518} {\bibfield  {journal} {\bibinfo  {journal} {Phys.
  Rev. B}\ }\textbf {\bibinfo {volume} {95}},\ \bibinfo {pages} {174518}
  (\bibinfo {year} {2017})}\BibitemShut {NoStop}%
\bibitem [{\citenamefont {Kashiwaya}\ \emph {et~al.}(2019)\citenamefont
  {Kashiwaya}, \citenamefont {Saitoh}, \citenamefont {Kashiwaya}, \citenamefont
  {Koyanagi}, \citenamefont {Sato}, \citenamefont {Yada}, \citenamefont
  {Tanaka},\ and\ \citenamefont {Maeno}}]{Kashiwaya2019}%
  \BibitemOpen
  \bibfield  {author} {\bibinfo {author} {\bibfnamefont {S.}~\bibnamefont
  {Kashiwaya}}, \bibinfo {author} {\bibfnamefont {K.}~\bibnamefont {Saitoh}},
  \bibinfo {author} {\bibfnamefont {H.}~\bibnamefont {Kashiwaya}}, \bibinfo
  {author} {\bibfnamefont {M.}~\bibnamefont {Koyanagi}}, \bibinfo {author}
  {\bibfnamefont {M.}~\bibnamefont {Sato}}, \bibinfo {author} {\bibfnamefont
  {K.}~\bibnamefont {Yada}}, \bibinfo {author} {\bibfnamefont {Y.}~\bibnamefont
  {Tanaka}}, \ and\ \bibinfo {author} {\bibfnamefont {Y.}~\bibnamefont
  {Maeno}},\ }\href {\doibase 10.1103/PhysRevB.100.094530} {\bibfield
  {journal} {\bibinfo  {journal} {Phys. Rev. B}\ }\textbf {\bibinfo {volume}
  {100}},\ \bibinfo {pages} {094530} (\bibinfo {year} {2019})}\BibitemShut
  {NoStop}%
\bibitem [{\citenamefont {Hassinger}\ \emph {et~al.}(2017)\citenamefont
  {Hassinger}, \citenamefont {Bourgeois-Hope}, \citenamefont {Taniguchi},
  \citenamefont {Ren\'e~de Cotret}, \citenamefont {Grissonnanche},
  \citenamefont {Anwar}, \citenamefont {Maeno}, \citenamefont
  {Doiron-Leyraud},\ and\ \citenamefont {Taillefer}}]{Hassinger_PRX_2017}%
  \BibitemOpen
  \bibfield  {author} {\bibinfo {author} {\bibfnamefont {E.}~\bibnamefont
  {Hassinger}}, \bibinfo {author} {\bibfnamefont {P.}~\bibnamefont
  {Bourgeois-Hope}}, \bibinfo {author} {\bibfnamefont {H.}~\bibnamefont
  {Taniguchi}}, \bibinfo {author} {\bibfnamefont {S.}~\bibnamefont {Ren\'e~de
  Cotret}}, \bibinfo {author} {\bibfnamefont {G.}~\bibnamefont
  {Grissonnanche}}, \bibinfo {author} {\bibfnamefont {M.~S.}\ \bibnamefont
  {Anwar}}, \bibinfo {author} {\bibfnamefont {Y.}~\bibnamefont {Maeno}},
  \bibinfo {author} {\bibfnamefont {N.}~\bibnamefont {Doiron-Leyraud}}, \ and\
  \bibinfo {author} {\bibfnamefont {L.}~\bibnamefont {Taillefer}},\ }\href
  {\doibase 10.1103/PhysRevX.7.011032} {\bibfield  {journal} {\bibinfo
  {journal} {Phys. Rev. X}\ }\textbf {\bibinfo {volume} {7}},\ \bibinfo {pages}
  {011032} (\bibinfo {year} {2017})}\BibitemShut {NoStop}%
\bibitem [{\citenamefont {Kittaka}\ \emph {et~al.}(2018)\citenamefont
  {Kittaka}, \citenamefont {Nakamura}, \citenamefont {Sakakibara},
  \citenamefont {Kikugawa}, \citenamefont {Terashima}, \citenamefont {Uji},
  \citenamefont {Sokolov}, \citenamefont {Mackenzie}, \citenamefont {Irie},
  \citenamefont {Tsutsumi} \emph {et~al.}}]{Kittaka_JPSJ_2018}%
  \BibitemOpen
  \bibfield  {author} {\bibinfo {author} {\bibfnamefont {S.}~\bibnamefont
  {Kittaka}}, \bibinfo {author} {\bibfnamefont {S.}~\bibnamefont {Nakamura}},
  \bibinfo {author} {\bibfnamefont {T.}~\bibnamefont {Sakakibara}}, \bibinfo
  {author} {\bibfnamefont {N.}~\bibnamefont {Kikugawa}}, \bibinfo {author}
  {\bibfnamefont {T.}~\bibnamefont {Terashima}}, \bibinfo {author}
  {\bibfnamefont {S.}~\bibnamefont {Uji}}, \bibinfo {author} {\bibfnamefont
  {D.~A.}\ \bibnamefont {Sokolov}}, \bibinfo {author} {\bibfnamefont {A.~P.}\
  \bibnamefont {Mackenzie}}, \bibinfo {author} {\bibfnamefont {K.}~\bibnamefont
  {Irie}}, \bibinfo {author} {\bibfnamefont {Y.}~\bibnamefont {Tsutsumi}},
  \emph {et~al.},\ }\href@noop {} {\bibfield  {journal} {\bibinfo  {journal}
  {J. Phys. Soc. Jpn.}\ }\textbf {\bibinfo {volume} {87}},\ \bibinfo {pages}
  {093703} (\bibinfo {year} {2018})}\BibitemShut {NoStop}%
\bibitem [{\citenamefont {\ifmmode \check{Z}\else
  \v{Z}\fi{}uti\ifmmode~\acute{c}\else \'{c}\fi{}}\ and\ \citenamefont
  {Mazin}(2005)}]{Zutic_PRL_2005}%
  \BibitemOpen
  \bibfield  {author} {\bibinfo {author} {\bibfnamefont {I.}~\bibnamefont
  {\ifmmode \check{Z}\else \v{Z}\fi{}uti\ifmmode~\acute{c}\else \'{c}\fi{}}}\
  and\ \bibinfo {author} {\bibfnamefont {I.}~\bibnamefont {Mazin}},\ }\href
  {\doibase 10.1103/PhysRevLett.95.217004} {\bibfield  {journal} {\bibinfo
  {journal} {Phys. Rev. Lett.}\ }\textbf {\bibinfo {volume} {95}},\ \bibinfo
  {pages} {217004} (\bibinfo {year} {2005})}\BibitemShut {NoStop}%
\bibitem [{\citenamefont {Hara}\ and\ \citenamefont
  {Nagai}(1986)}]{Hara_PTP_1986}%
  \BibitemOpen
  \bibfield  {author} {\bibinfo {author} {\bibfnamefont {J.}~\bibnamefont
  {Hara}}\ and\ \bibinfo {author} {\bibfnamefont {K.}~\bibnamefont {Nagai}},\
  }\href {\doibase 10.1143/PTP.76.1237} {\bibfield  {journal} {\bibinfo
  {journal} {Progress of Theoretical Physics}\ }\textbf {\bibinfo {volume}
  {76}},\ \bibinfo {pages} {1237} (\bibinfo {year} {1986})}\BibitemShut
  {NoStop}%
\bibitem [{\citenamefont {Bruder}(1990)}]{Bruder}%
  \BibitemOpen
  \bibfield  {author} {\bibinfo {author} {\bibfnamefont {C.}~\bibnamefont
  {Bruder}},\ }\href {\doibase 10.1103/PhysRevB.41.4017} {\bibfield  {journal}
  {\bibinfo  {journal} {Phys. Rev. B}\ }\textbf {\bibinfo {volume} {41}},\
  \bibinfo {pages} {4017} (\bibinfo {year} {1990})}\BibitemShut {NoStop}%
\bibitem [{\citenamefont {Tanaka}\ and\ \citenamefont
  {Kashiwaya}(1995{\natexlab{a}})}]{TK95}%
  \BibitemOpen
  \bibfield  {author} {\bibinfo {author} {\bibfnamefont {Y.}~\bibnamefont
  {Tanaka}}\ and\ \bibinfo {author} {\bibfnamefont {S.}~\bibnamefont
  {Kashiwaya}},\ }\href@noop {} {\bibfield  {journal} {\bibinfo  {journal}
  {Phys. Rev. Lett.}\ }\textbf {\bibinfo {volume} {74}},\ \bibinfo {pages}
  {3451} (\bibinfo {year} {1995}{\natexlab{a}})}\BibitemShut {NoStop}%
\bibitem [{\citenamefont {Buchholtz}\ and\ \citenamefont
  {Zwicknagl}(1981)}]{ABS}%
  \BibitemOpen
  \bibfield  {author} {\bibinfo {author} {\bibfnamefont {L.~J.}\ \bibnamefont
  {Buchholtz}}\ and\ \bibinfo {author} {\bibfnamefont {G.}~\bibnamefont
  {Zwicknagl}},\ }\href@noop {} {\bibfield  {journal} {\bibinfo  {journal}
  {Phys. Rev. B}\ }\textbf {\bibinfo {volume} {23}},\ \bibinfo {pages} {5788}
  (\bibinfo {year} {1981})}\BibitemShut {NoStop}%
\bibitem [{\citenamefont {Yamashiro}\ \emph {et~al.}(1997)\citenamefont
  {Yamashiro}, \citenamefont {Tanaka},\ and\ \citenamefont
  {Kashiwaya}}]{Yamashiro97}%
  \BibitemOpen
  \bibfield  {author} {\bibinfo {author} {\bibfnamefont {M.}~\bibnamefont
  {Yamashiro}}, \bibinfo {author} {\bibfnamefont {Y.}~\bibnamefont {Tanaka}}, \
  and\ \bibinfo {author} {\bibfnamefont {S.}~\bibnamefont {Kashiwaya}},\
  }\href@noop {} {\bibfield  {journal} {\bibinfo  {journal} {Phys. Rev. B}\
  }\textbf {\bibinfo {volume} {56}},\ \bibinfo {pages} {7847} (\bibinfo {year}
  {1997})}\BibitemShut {NoStop}%
\bibitem [{\citenamefont {Honerkamp}\ and\ \citenamefont
  {Sigrist}(1998)}]{Honerkamp98}%
  \BibitemOpen
  \bibfield  {author} {\bibinfo {author} {\bibfnamefont {C.}~\bibnamefont
  {Honerkamp}}\ and\ \bibinfo {author} {\bibfnamefont {M.}~\bibnamefont
  {Sigrist}},\ }\href@noop {} {\bibfield  {journal} {\bibinfo  {journal} {J.
  Low Temp. Phys.}\ }\textbf {\bibinfo {volume} {111}},\ \bibinfo {pages} {895}
  (\bibinfo {year} {1998})}\BibitemShut {NoStop}%
\bibitem [{\citenamefont {Hu}(1994)}]{Hu}%
  \BibitemOpen
  \bibfield  {author} {\bibinfo {author} {\bibfnamefont {C.~R.}\ \bibnamefont
  {Hu}},\ }\href@noop {} {\bibfield  {journal} {\bibinfo  {journal} {Phys. Rev.
  Lett.}\ }\textbf {\bibinfo {volume} {72}},\ \bibinfo {pages} {1526} (\bibinfo
  {year} {1994})}\BibitemShut {NoStop}%
\bibitem [{\citenamefont {Kashiwaya}\ and\ \citenamefont
  {Tanaka}(2000)}]{ABSR1}%
  \BibitemOpen
  \bibfield  {author} {\bibinfo {author} {\bibfnamefont {S.}~\bibnamefont
  {Kashiwaya}}\ and\ \bibinfo {author} {\bibfnamefont {Y.}~\bibnamefont
  {Tanaka}},\ }\href@noop {} {\bibfield  {journal} {\bibinfo  {journal} {Rep.
  Prog. Phys.}\ }\textbf {\bibinfo {volume} {63}},\ \bibinfo {pages} {1641}
  (\bibinfo {year} {2000})}\BibitemShut {NoStop}%
\bibitem [{\citenamefont {L{\"o}fwander}\ \emph {et~al.}(2001)\citenamefont
  {L{\"o}fwander}, \citenamefont {Shumeiko},\ and\ \citenamefont
  {Wendin}}]{ABSR2}%
  \BibitemOpen
  \bibfield  {author} {\bibinfo {author} {\bibfnamefont {T.}~\bibnamefont
  {L{\"o}fwander}}, \bibinfo {author} {\bibfnamefont {V.~S.}\ \bibnamefont
  {Shumeiko}}, \ and\ \bibinfo {author} {\bibfnamefont {G.}~\bibnamefont
  {Wendin}},\ }\href@noop {} {\bibfield  {journal} {\bibinfo  {journal}
  {Supercond. Sci. Technol.}\ }\textbf {\bibinfo {volume} {14}},\ \bibinfo
  {pages} {R53} (\bibinfo {year} {2001})}\BibitemShut {NoStop}%
\bibitem [{\citenamefont {Asano}\ \emph {et~al.}(2004)\citenamefont {Asano},
  \citenamefont {Tanaka},\ and\ \citenamefont {Kashiwaya}}]{Asano_PRB_2004}%
  \BibitemOpen
  \bibfield  {author} {\bibinfo {author} {\bibfnamefont {Y.}~\bibnamefont
  {Asano}}, \bibinfo {author} {\bibfnamefont {Y.}~\bibnamefont {Tanaka}}, \
  and\ \bibinfo {author} {\bibfnamefont {S.}~\bibnamefont {Kashiwaya}},\ }\href
  {\doibase 10.1103/PhysRevB.69.134501} {\bibfield  {journal} {\bibinfo
  {journal} {Phys. Rev. B}\ }\textbf {\bibinfo {volume} {69}},\ \bibinfo
  {pages} {134501} (\bibinfo {year} {2004})}\BibitemShut {NoStop}%
\bibitem [{\citenamefont {Suderow}\ \emph {et~al.}(2009)\citenamefont
  {Suderow}, \citenamefont {Crespo}, \citenamefont {Guillamon}, \citenamefont
  {Vieira}, \citenamefont {Servant}, \citenamefont {Lejay}, \citenamefont
  {Brison},\ and\ \citenamefont {Flouquet}}]{Suderow_2009}%
  \BibitemOpen
  \bibfield  {author} {\bibinfo {author} {\bibfnamefont {H.}~\bibnamefont
  {Suderow}}, \bibinfo {author} {\bibfnamefont {V.}~\bibnamefont {Crespo}},
  \bibinfo {author} {\bibfnamefont {I.}~\bibnamefont {Guillamon}}, \bibinfo
  {author} {\bibfnamefont {S.}~\bibnamefont {Vieira}}, \bibinfo {author}
  {\bibfnamefont {F.}~\bibnamefont {Servant}}, \bibinfo {author} {\bibfnamefont
  {P.}~\bibnamefont {Lejay}}, \bibinfo {author} {\bibfnamefont {J.~P.}\
  \bibnamefont {Brison}}, \ and\ \bibinfo {author} {\bibfnamefont
  {J.}~\bibnamefont {Flouquet}},\ }\href {\doibase
  10.1088/1367-2630/11/9/093004} {\bibfield  {journal} {\bibinfo  {journal}
  {New Journal of Physics}\ }\textbf {\bibinfo {volume} {11}},\ \bibinfo
  {pages} {093004} (\bibinfo {year} {2009})}\BibitemShut {NoStop}%
\bibitem [{\citenamefont {Firmo}\ \emph {et~al.}(2013)\citenamefont {Firmo},
  \citenamefont {Lederer}, \citenamefont {Lupien}, \citenamefont {Mackenzie},
  \citenamefont {Davis},\ and\ \citenamefont {Kivelson}}]{Firmo_PRB_2013}%
  \BibitemOpen
  \bibfield  {author} {\bibinfo {author} {\bibfnamefont {I.~A.}\ \bibnamefont
  {Firmo}}, \bibinfo {author} {\bibfnamefont {S.}~\bibnamefont {Lederer}},
  \bibinfo {author} {\bibfnamefont {C.}~\bibnamefont {Lupien}}, \bibinfo
  {author} {\bibfnamefont {A.~P.}\ \bibnamefont {Mackenzie}}, \bibinfo {author}
  {\bibfnamefont {J.~C.}\ \bibnamefont {Davis}}, \ and\ \bibinfo {author}
  {\bibfnamefont {S.~A.}\ \bibnamefont {Kivelson}},\ }\href {\doibase
  10.1103/PhysRevB.88.134521} {\bibfield  {journal} {\bibinfo  {journal} {Phys.
  Rev. B}\ }\textbf {\bibinfo {volume} {88}},\ \bibinfo {pages} {134521}
  (\bibinfo {year} {2013})}\BibitemShut {NoStop}%
\bibitem [{\citenamefont {Sharma}\ \emph {et~al.}(2019)\citenamefont {Sharma},
  \citenamefont {Edkins}, \citenamefont {Wang}, \citenamefont {Kostin},
  \citenamefont {Sow}, \citenamefont {Maeno}, \citenamefont {Mackenzie},
  \citenamefont {Davis},\ and\ \citenamefont {Madhavan}}]{Madhavan}%
  \BibitemOpen
  \bibfield  {author} {\bibinfo {author} {\bibfnamefont {R.}~\bibnamefont
  {Sharma}}, \bibinfo {author} {\bibfnamefont {S.~D.}\ \bibnamefont {Edkins}},
  \bibinfo {author} {\bibfnamefont {Z.}~\bibnamefont {Wang}}, \bibinfo {author}
  {\bibfnamefont {A.}~\bibnamefont {Kostin}}, \bibinfo {author} {\bibfnamefont
  {C.}~\bibnamefont {Sow}}, \bibinfo {author} {\bibfnamefont {Y.}~\bibnamefont
  {Maeno}}, \bibinfo {author} {\bibfnamefont {A.~P.}\ \bibnamefont
  {Mackenzie}}, \bibinfo {author} {\bibfnamefont {J.}~\bibnamefont {Davis}}, \
  and\ \bibinfo {author} {\bibfnamefont {V.}~\bibnamefont {Madhavan}},\
  }\href@noop {} {\bibfield  {journal} {\bibinfo  {journal} {arXiv preprint
  arXiv:1912.02798}\ } (\bibinfo {year} {2019})}\BibitemShut {NoStop}%
\bibitem [{Note1()}]{Note1}%
  \BibitemOpen
  \bibinfo {note} {In this paper, we consider subdominant component only for
  the FW pairing. Subdominant components in a spin-triplet superconductor can
  change the direction of the $d$-vector, resulting in the mixing between spin
  subspace. Such spin mixing can affects the surface ABSs.}\BibitemShut {Stop}%
\bibitem [{\citenamefont {Blonder}\ \emph {et~al.}(1982)\citenamefont
  {Blonder}, \citenamefont {Tinkham},\ and\ \citenamefont {Klapwijk}}]{BTK}%
  \BibitemOpen
  \bibfield  {author} {\bibinfo {author} {\bibfnamefont {G.~E.}\ \bibnamefont
  {Blonder}}, \bibinfo {author} {\bibfnamefont {M.}~\bibnamefont {Tinkham}}, \
  and\ \bibinfo {author} {\bibfnamefont {T.~M.}\ \bibnamefont {Klapwijk}},\
  }\href {\doibase 10.1103/PhysRevB.25.4515} {\bibfield  {journal} {\bibinfo
  {journal} {Phys. Rev. B}\ }\textbf {\bibinfo {volume} {25}},\ \bibinfo
  {pages} {4515} (\bibinfo {year} {1982})}\BibitemShut {NoStop}%
\bibitem [{\citenamefont {Yamashiro}\ \emph
  {et~al.}(1998{\natexlab{b}})\citenamefont {Yamashiro}, \citenamefont
  {Tanaka}, \citenamefont {Tanuma},\ and\ \citenamefont {Kashiwaya}}]{YTK98}%
  \BibitemOpen
  \bibfield  {author} {\bibinfo {author} {\bibfnamefont {M.}~\bibnamefont
  {Yamashiro}}, \bibinfo {author} {\bibfnamefont {Y.}~\bibnamefont {Tanaka}},
  \bibinfo {author} {\bibfnamefont {Y.}~\bibnamefont {Tanuma}}, \ and\ \bibinfo
  {author} {\bibfnamefont {S.}~\bibnamefont {Kashiwaya}},\ }\href@noop {}
  {\bibfield  {journal} {\bibinfo  {journal} {J. Phys. Soc. Jpn.}\ }\textbf
  {\bibinfo {volume} {67}},\ \bibinfo {pages} {3224} (\bibinfo {year}
  {1998}{\natexlab{b}})}\BibitemShut {NoStop}%
\bibitem [{\citenamefont {Kwon}\ \emph {et~al.}(2004)\citenamefont {Kwon},
  \citenamefont {Sengupta},\ and\ \citenamefont {Yakovenko}}]{Yakovenko}%
  \BibitemOpen
  \bibfield  {author} {\bibinfo {author} {\bibfnamefont {H.}~\bibnamefont
  {Kwon}}, \bibinfo {author} {\bibfnamefont {K.}~\bibnamefont {Sengupta}}, \
  and\ \bibinfo {author} {\bibfnamefont {V.}~\bibnamefont {Yakovenko}},\
  }\href@noop {} {\bibfield  {journal} {\bibinfo  {journal} {Eur. Phys. J. B}\
  }\textbf {\bibinfo {volume} {37}},\ \bibinfo {pages} {349} (\bibinfo {year}
  {2004})}\BibitemShut {NoStop}%
\bibitem [{Note2()}]{Note2}%
  \BibitemOpen
  \bibinfo {note} {The ABSs are formed by the interference between the
  quasiparticles propagating with $k_z$ and $-k_z$. The ABSs are present when
  the phases of the pair potentials $\Delta (\protect \boldsymbol {k}_\parallel
  , k_z)$ and $\Delta (\protect \boldsymbol {k}_\parallel ,- k_z)$ are
  different. In particular, the ABS appears at the zero energy when the
  condition $\phi (\protect \boldsymbol {k}_\parallel , k_z) -\phi (\protect
  \boldsymbol {k}_\parallel , k_z)= \pi $ is satisfied, where $\phi $ is the
  phase of the pair potential \cite {Hu,TK95,ABSR1,ABSR2}}\BibitemShut
  {NoStop}%
\bibitem [{Note3()}]{Note3}%
  \BibitemOpen
  \bibinfo {note} {The energy scale depends also on the size of the FS in the N
  region}\BibitemShut {NoStop}%
\bibitem [{\citenamefont {Kobayashi}\ \emph {et~al.}(2015)\citenamefont
  {Kobayashi}, \citenamefont {Tanaka},\ and\ \citenamefont
  {Sato}}]{Kobayashi_PRB_2015}%
  \BibitemOpen
  \bibfield  {author} {\bibinfo {author} {\bibfnamefont {S.}~\bibnamefont
  {Kobayashi}}, \bibinfo {author} {\bibfnamefont {Y.}~\bibnamefont {Tanaka}}, \
  and\ \bibinfo {author} {\bibfnamefont {M.}~\bibnamefont {Sato}},\ }\href
  {\doibase 10.1103/PhysRevB.92.214514} {\bibfield  {journal} {\bibinfo
  {journal} {Phys. Rev. B}\ }\textbf {\bibinfo {volume} {92}},\ \bibinfo
  {pages} {214514} (\bibinfo {year} {2015})}\BibitemShut {NoStop}%
\bibitem [{Note4()}]{Note4}%
  \BibitemOpen
  \bibinfo {note} {In the square-lattice tight-binding model, $G_{\protect
  \mathrm {NS}}$ depends on the direction reflecting the FS anisotropy in the
  $ab$ plane}\BibitemShut {NoStop}%
\bibitem [{\citenamefont {Tanaka}\ and\ \citenamefont
  {Kashiwaya}(1995{\natexlab{b}})}]{Tanaka_PRL_1995}%
  \BibitemOpen
  \bibfield  {author} {\bibinfo {author} {\bibfnamefont {Y.}~\bibnamefont
  {Tanaka}}\ and\ \bibinfo {author} {\bibfnamefont {S.}~\bibnamefont
  {Kashiwaya}},\ }\href {\doibase 10.1103/PhysRevLett.74.3451} {\bibfield
  {journal} {\bibinfo  {journal} {Phys. Rev. Lett.}\ }\textbf {\bibinfo
  {volume} {74}},\ \bibinfo {pages} {3451} (\bibinfo {year}
  {1995}{\natexlab{b}})}\BibitemShut {NoStop}%
\bibitem [{\citenamefont {Alff}\ \emph {et~al.}(1997)\citenamefont {Alff},
  \citenamefont {Takashima}, \citenamefont {Kashiwaya}, \citenamefont {Terada},
  \citenamefont {Ihara}, \citenamefont {Tanaka}, \citenamefont {Koyanagi},\
  and\ \citenamefont {Kajimura}}]{Experiment4}%
  \BibitemOpen
  \bibfield  {author} {\bibinfo {author} {\bibfnamefont {L.}~\bibnamefont
  {Alff}}, \bibinfo {author} {\bibfnamefont {H.}~\bibnamefont {Takashima}},
  \bibinfo {author} {\bibfnamefont {S.}~\bibnamefont {Kashiwaya}}, \bibinfo
  {author} {\bibfnamefont {N.}~\bibnamefont {Terada}}, \bibinfo {author}
  {\bibfnamefont {H.}~\bibnamefont {Ihara}}, \bibinfo {author} {\bibfnamefont
  {Y.}~\bibnamefont {Tanaka}}, \bibinfo {author} {\bibfnamefont
  {M.}~\bibnamefont {Koyanagi}}, \ and\ \bibinfo {author} {\bibfnamefont
  {K.}~\bibnamefont {Kajimura}},\ }\href@noop {} {\bibfield  {journal}
  {\bibinfo  {journal} {Phys. Rev. B}\ }\textbf {\bibinfo {volume} {55}},\
  \bibinfo {pages} {R14757} (\bibinfo {year} {1997})}\BibitemShut {NoStop}%
\bibitem [{\citenamefont {Wei}\ \emph {et~al.}(1998)\citenamefont {Wei},
  \citenamefont {Yeh}, \citenamefont {Garrigus},\ and\ \citenamefont
  {Strasik}}]{Experiment5}%
  \BibitemOpen
  \bibfield  {author} {\bibinfo {author} {\bibfnamefont {J.~Y.~T.}\
  \bibnamefont {Wei}}, \bibinfo {author} {\bibfnamefont {N.-C.}\ \bibnamefont
  {Yeh}}, \bibinfo {author} {\bibfnamefont {D.~F.}\ \bibnamefont {Garrigus}}, \
  and\ \bibinfo {author} {\bibfnamefont {M.}~\bibnamefont {Strasik}},\
  }\href@noop {} {\bibfield  {journal} {\bibinfo  {journal} {Phys. Rev. Lett.}\
  }\textbf {\bibinfo {volume} {81}},\ \bibinfo {pages} {2542} (\bibinfo {year}
  {1998})}\BibitemShut {NoStop}%
\bibitem [{\citenamefont {Yada}\ \emph {et~al.}(2014)\citenamefont {Yada},
  \citenamefont {Golubov}, \citenamefont {Tanaka},\ and\ \citenamefont
  {Kashiwaya}}]{Yada_JPSJ_2014}%
  \BibitemOpen
  \bibfield  {author} {\bibinfo {author} {\bibfnamefont {K.}~\bibnamefont
  {Yada}}, \bibinfo {author} {\bibfnamefont {A.~A.}\ \bibnamefont {Golubov}},
  \bibinfo {author} {\bibfnamefont {Y.}~\bibnamefont {Tanaka}}, \ and\ \bibinfo
  {author} {\bibfnamefont {S.}~\bibnamefont {Kashiwaya}},\ }\href
  {https://doi.org/10.7566/JPSJ.83.074706} {\bibfield  {journal} {\bibinfo
  {journal} {J. Phys. Soc. Jpn.}\ }\textbf {\bibinfo {volume} {83}},\ \bibinfo
  {pages} {074706} (\bibinfo {year} {2014})}\BibitemShut {NoStop}%
\bibitem [{\citenamefont {Barash}\ \emph {et~al.}(1997)\citenamefont {Barash},
  \citenamefont {Svidzinsky},\ and\ \citenamefont {Burkhardt}}]{Barash97}%
  \BibitemOpen
  \bibfield  {author} {\bibinfo {author} {\bibfnamefont {Y.}~\bibnamefont
  {Barash}}, \bibinfo {author} {\bibfnamefont {A.}~\bibnamefont {Svidzinsky}},
  \ and\ \bibinfo {author} {\bibfnamefont {H.}~\bibnamefont {Burkhardt}},\
  }\href@noop {} {\bibfield  {journal} {\bibinfo  {journal} {Phys. Rev. B}\
  }\textbf {\bibinfo {volume} {55}},\ \bibinfo {pages} {15282} (\bibinfo {year}
  {1997})}\BibitemShut {NoStop}%
\bibitem [{\citenamefont {Nagato}\ \emph {et~al.}(1993)\citenamefont {Nagato},
  \citenamefont {Nagai},\ and\ \citenamefont {Hara}}]{Nagato_JLTP_1993}%
  \BibitemOpen
  \bibfield  {author} {\bibinfo {author} {\bibfnamefont {Y.}~\bibnamefont
  {Nagato}}, \bibinfo {author} {\bibfnamefont {K.}~\bibnamefont {Nagai}}, \
  and\ \bibinfo {author} {\bibfnamefont {J.}~\bibnamefont {Hara}},\ }\href
  {https://doi.org/10.1007/BF00682280} {\bibfield  {journal} {\bibinfo
  {journal} {J. Low Temp. Phys.}\ }\textbf {\bibinfo {volume} {93}},\ \bibinfo
  {pages} {33} (\bibinfo {year} {1993})}\BibitemShut {NoStop}%
\bibitem [{\citenamefont {Nagato}\ \emph {et~al.}(1998)\citenamefont {Nagato},
  \citenamefont {Yamamoto},\ and\ \citenamefont {Nagai}}]{Nagato_JLTP_1998}%
  \BibitemOpen
  \bibfield  {author} {\bibinfo {author} {\bibfnamefont {Y.}~\bibnamefont
  {Nagato}}, \bibinfo {author} {\bibfnamefont {M.}~\bibnamefont {Yamamoto}}, \
  and\ \bibinfo {author} {\bibfnamefont {K.}~\bibnamefont {Nagai}},\
  }\href@noop {} {\bibfield  {journal} {\bibinfo  {journal} {J. Low Temp.
  Phys.}\ }\textbf {\bibinfo {volume} {110}},\ \bibinfo {pages} {1135}
  (\bibinfo {year} {1998})}\BibitemShut {NoStop}%
\bibitem [{\citenamefont {Tanuma}\ \emph {et~al.}(2001)\citenamefont {Tanuma},
  \citenamefont {Tanaka},\ and\ \citenamefont {Kashiwaya}}]{Tanuma01}%
  \BibitemOpen
  \bibfield  {author} {\bibinfo {author} {\bibfnamefont {Y.}~\bibnamefont
  {Tanuma}}, \bibinfo {author} {\bibfnamefont {Y.}~\bibnamefont {Tanaka}}, \
  and\ \bibinfo {author} {\bibfnamefont {S.}~\bibnamefont {Kashiwaya}},\
  }\href@noop {} {\bibfield  {journal} {\bibinfo  {journal} {Phys. Rev. B}\
  }\textbf {\bibinfo {volume} {64}},\ \bibinfo {pages} {214519} (\bibinfo
  {year} {2001})}\BibitemShut {NoStop}%
\end{thebibliography}%

\end{document}